\documentclass[aps,preprint,tightenlines,showkeys,nofootinbib]{revtex4}

\usepackage{graphicx}  
\usepackage{bm}  
\usepackage{amsmath}
\newcommand{\beq}{\begin{equation}}
\newcommand{\eeq}{\end{equation}}
\newcommand{\bea}{\begin{eqnarray}}
\newcommand{\eea}{\end{eqnarray}} 
\newcommand{\beqa}{\begin{eqnarray}}
\newcommand{\eeqa}{\end{eqnarray}}

\newcommand{\simlt}{\stackrel{<}{{}_\sim}}

\begin{document}
\title{Range Corrections to Three-Body Observables near a Feshbach Resonance}
\author{L. Platter}\email{lplatter@mps.ohio-state.edu}
\affiliation{Department of Physics and Astronomy,
Ohio University, Athens, OH 45701, USA\\}
\affiliation{\vspace*{-0.4cm}Department of Physics,
The Ohio State University, Columbus, OH\ 43210, USA\\}
\author{C. Ji}\email{jichen@phy.ohiou.edu}
\affiliation{Department of Physics and Astronomy,
Ohio University, Athens, OH 45701, USA\\}
\author{D. R. Phillips}\email{phillips@phy.ohiou.edu}
\affiliation{Department of Physics and Astronomy,
Ohio University, Athens, OH 45701, USA\\}

\date{\today}
\begin{abstract}
A non-relativistic system of three identical particles will display a
rich set of universal features known as {\it Efimov} physics if the
scattering length $a$ is much larger than the range $l$ of the
underlying two-body interaction. An appropriate effective theory
facilitates the derivation of both results in the $|a| \rightarrow
\infty$ limit and finite-$l/a$ corrections to observables of
interest. Here we use such an effective-theory treatment to
consider the impact of corrections linear in the two-body effective
range, $r_s$ on the three-boson bound-state spectrum and recombination rate
for $|a| \gg |r_s|$. We do this by first deriving results appropriate to
the strict limit $|a| \rightarrow \infty$ in coordinate space. We
then extend these results to finite $a$ using once-subtracted
momentum-space integral equations.  We also discuss the implications
of our results for experiments that probe three-body recombination in
Bose-Einstein condensates near a Feshbach resonance.
\end{abstract}


\maketitle
\section{Introduction}
\label{sec:intro}
Systems in which the  two-body scattering length, $a$, is much
larger than the range, $l$, of the underlying interaction are said
to be close to the ``unitary limit", $|a| \rightarrow \infty$. In that 
limit the
two-body scattering saturates the bound set by the requirement
that the S-matrix be unitary. The dynamics of such systems for energies
$E \sim \hbar^2/(M a^2)$ ($M$ is the particle mass)  is universal,
in the sense that it is independent of any details of the underlying two-body interaction.
One obvious example of this ``universality" is that for $a>0$ all such systems
have a two-body bound state ({\it dimer}) with binding energy $E_D \approx \hbar^2/(M a^2)$.

A non-relativistic system of three identical particles 
with $|a| \gg l$ displays a rich
set of universal features \cite{Braaten:2004rn}.
In the limit $|a| \rightarrow \infty$ the system has an infinite
tower of three-body bound states ({\it trimers}).
Their binding energies are:
\beq
B_n=(e^{-2\pi s_0})^{n-n_*} B_{n*}
\label{eq:efispect}
\eeq
where $B_{n*}$ denotes the energy of an arbitrary state in the
spectrum, and the numerical constant $s_0 \approx 1.00624$. The
nature of the geometric spectrum (\ref{eq:efispect}) is a consequence of the
discrete scale invariance of the three-body problem in the joint limit $|a|
\rightarrow \infty$, $l \rightarrow 0$.  Efimov, who made this
discovery, pointed out that this discrete scale invariance is also
relevant for finite $a$ and that it should be observable in systems
for which $l/|a| \neq 0$, but is still $\ll 1$ \cite{Efi71,Efi79}.

Evidence for Efimov physics in cold atoms was recently presented by
the Innsbruck group \cite{Grimm06}. Using a magnetic field to control the scattering
length, they measured the recombination rate of cold $^{133}$Cs atoms
in the lowest hyperfine state and found resonant enhancement at $a\sim
-850 a_0$, which can be attributed to the existence of an Efimov trimer
at the three-body threshold. Approaches which employ a zero-range model and therefore 
account for the separation of scales between $a$ and $l$ by taking the limit 
$l \rightarrow 0$ are very
successful in describing this experimental data~\cite{Braaten:2003yc,Braaten:2008kx}. 
But a systematic
calculation of the corrections to these results due to finite-range effects
is desirable, in order to identify the limitations of these
previous calculations and also to extend the range of applicability of
approaches based on the scale hierarchy $l \ll |a|$.

In recent years Efimov's results have been rederived in the framework
of an effective field theory (EFT) \cite{Bedaque:1998kg}. This EFT employs the ratio $l/|a|$ as
a small expansion parameter and allows for a systematically improvable
and model-independent calculation of observables of few-body systems
with a large scattering length.
The leading order (LO) of this EFT  
reproduces all universal features previously discovered by
Efimov.

In this paper we will exploit effective theory methods to derive 
the correction linear in the effective range to this universal
behavior (\ref{eq:efispect}). In the EFT expansion this corresponds to the
next-to-leading order (NLO) piece. The value of the two-body effective range,
$r_s$, sets the size of these corrections. We evaluate their impact
perturbatively, and so obtain for the three-body bound-state spectrum a result of
the form:
\beq
B_n=B_{n*}
\left[F_n\left(\frac{\gamma}{\kappa_*}\right)
+ \kappa_* \,r_s \,G_n\left(\frac{\gamma}{\kappa_*}\right) + O[(\kappa_* \, r_s)^2]\right],
\label{eq:Gn1}
\eeq
where $\kappa_*=\sqrt{M B_{n*}}$ is the binding momentum of the $n*$th
state in the three-body bound-state spectrum, $\gamma=1/a$ (at LO)
is the binding momentum of the two-body bound state
that is present for $a > 0$, and the functions $F_n$
obey~\cite{Braaten:2004rn}
\beq
F_n(e^{\frac{m \pi}{s_0}} x)=e^{-\frac{2 m \pi}{s_0}} F_{n-m}(x).
\label{eq:Fnreln}
\eeq
We will show that the functions $G_n$ obey approximate relations similar to
Eq.~(\ref{eq:Fnreln}), since discrete scale invariance connects the
 $G_n$'s for different values of $n$.  Knowledge of the
functions $F_{n*}$ and $G_{n}$ therefore provides information on the
behavior of the bound-state spectrum in the vicinity of a narrow
Feshbach resonance to a precision given by $(r_s/a)^2$. In this work we
will give results for the function $G_{n*}$  from the unitary limit
up to the value of $a$ where the three-body bound state under consideration
hits the threshold. In particular, we will show that
this function is zero at particular values of $a$.

We do this by formally studying systems in which the effective
range, $r_s$, is significantly larger than the range $l$ of the
underlying two-body potential, i.e. the scale hierarchy is $l \ll |r_s|
\ll |a|$. It is important to point out that in such a three-scale problem it is possible to be 
in the unitary limit, but still have range effects present in the result, i.e. to not have achieved what is sometimes called
the ``scaling limit", where $r_s \rightarrow 0$ and $l \rightarrow 0$. In fact,
in the EFT formulation of the problem the range of the
underlying two-body potential, $l$, provides a limiting distance,
because the EFT does not specify any details of the two-body
potential, and so it does not accurately describe the three-body
dynamics in those regions of configuration space where any two-body
distance is less than $l$.  Hence it is natural to place a cutoff
$\Lambda$ on the momentum-space integrals present in the EFT
calculations, and choose $\Lambda \sim 1/l$. In our case we are
interested in systems with $l \ll |r_s|$, and so we will ultimately take
the limit $\Lambda \rightarrow \infty \equiv l \rightarrow 0$ in order to obtain our
results. The effective range, $r_s$, will, however, be kept finite.

The inclusion and effects of range corrections in the EFT with contact
interactions have been studied extensively. In \cite{Hammer:2001gh}
range corrections in the three-nucleon system were calculated
perturbatively in momentum space. A formalism which employs a
partial resummation of range effects was developed in Refs.~\cite{Bedaque:2002yg,Platter:2006a,Platter:2006b} and observables
were calculated up to next-to-next-to-leading order (N2LO) in the
$l/a$ expansion. Hyperradial coordinates and the Wilsonian renormalization group were used to re-derive
the leading-order amplitude and develop a power counting for sub-leading effects in Ref.~\cite{Barford:2004fz}.

In Sec.~\ref{sec-rspace} we will briefly review the hyperradial
formalism and show analytically that in the limit of infinite
scattering length no linear range correction exists, i.e. that
$G_n(0)=0$ for all $n$. In Sec.~\ref{sec-pspace} we will review a formalism
that allows us to compute
numerical results for this linear correction away from the unitary
limit and present numerical results for the three-body bound-state spectrum and the 
function $G_n$.  In Sec.~\ref{sec-recombination} we will lay out how the formalism
developed for the analysis of Sec.~\ref{sec-pspace} can be extended to 
the computation of range corrections for the three-body recombination rate in a
cold gas of identical bosons. We will end with a summary and
discussion of our findings.

\section{Near a Feshbach resonance: results in the unitary limit, with
  non-zero effective range}

\label{sec-rspace}

In this section we first review the use of hyperradial coordinates for
the three-body problem in the unitary limit (Subsection~\ref{sec-hyperradrev}). 
Our presentation closely follows that of Ref.~\cite{Braaten:2004rn}. In order to
fix notation and methodology for the subsequent subsections
we derive the potential
between the single particle and the pair in hyperspherical coordinates, and
show that it behaves like one over the hyperradius, $R$, squared for
$|r_s| \ll R \ll |a|$. We explain why---in the joint limits $r_s
\rightarrow 0$ and $|a| \rightarrow \infty$---this results in the
Efimov spectrum described above, and compute the wave functions in
this limit. A somewhat different explanation of these results---one based on 
Efimov's original work~\cite{Efi71,Efi79}---can be found in Appendix A 
of Ref.~\cite{Barford:2004fz}.

We then show how to compute perturbative changes to this spectrum as
we move away from the scaling limit. In
Subsection~\ref{sec-linearcorrn}, we derive the terms in the
hyperspherical potential that scale as $r_s/R^3$, and
compute the matrix elements of this potential between the wave
functions obtained with the pure $1/R^2$ potential. These matrix
elements diverge in the ultraviolet, but we show that if the spectrum
is renormalized by demanding that short-distance physics at $R \sim r_s$
remains unchanged when we move from LO to NLO then there is no change
in the Efimov spectrum at $O(r_s)$. 

\subsection{Hyperradial formalism: a review}

\label{sec-hyperradrev}

The material of this subsection is based on the discussion of the
hyperradial formalism in the recent review of
Ref.~\cite{Braaten:2004rn}.   
For three particles of equal mass the
Jacobi coordinates are defined as:
\begin{equation}
{\bf r}_{ij}={\bf r}_i - {\bf r}_j~;\quad{\bf r}_{k,ij}={\bf r}_k -
\frac{1}{2}({\bf r}_i + {\bf r}_j)~,
\label{eq:Jacobi}
\end{equation}
where the triple $(ijk)$ is a cyclic permutation of the particle
indices $(123)$. 
The hyperradius $R$ and hyperangle $\alpha_k$ are then defined by
\begin{equation}
R^2=\frac{1}{3}({\bf r}_{12}^2 + {\bf r}_{23}^2 + {\bf
  r}_{31}^2)=\frac{1}{2} {\bf r}_{ij}^2 + \frac{2}{3}{\bf r}_{k,ij}^2; \quad
\alpha_k=\arctan\left(\frac{\sqrt{3} |{\bf r}_{ij}|}{2 |{\bf
    r}_{k,ij}|}\right).
\end{equation}
In the center-of-mass system the Schr\"odinger equation in hyperspherical
coordinates is given by
\begin{equation}
\left(T_R + T_{\alpha_k} + \frac{\Lambda_{k,ij}^2}{2 M R^2} +
V(R,\Omega)\right)
\Psi(R,\alpha,\Omega)=E \Psi(R,\alpha,\Omega),
\label{eq:hyperschro}
\end{equation}
with
\begin{eqnarray}
T_R&=&\frac{\hbar^2}{2 M} R^{-5/2} \left(-\frac{\partial}{\partial R^2} +
\frac{15}{4 R^2}\right) R^{5/2},\\
T_\alpha&=&\frac{\hbar^2}{2 M R^2} \frac{1}{\sin 2 \alpha}
\left(-\frac{\partial^2}{\partial \alpha^2} - 4\right) \sin 2 \alpha,\\
{\Lambda_{k,ij}}^2&=&\frac{\mathbf L_{ij}^2}{\sin^2\alpha_k}  +
\frac{\mathbf L_{k,ij}^2}{\cos^2 \alpha_k},
\label{eq:angmom}
\end{eqnarray}
where $\Omega=(\theta_{ij},\phi_{ij},\theta_{k,ij},\phi_{k,ij})$ and
the $L$s that appear in Eq.~(\ref{eq:angmom}) are the usual
angular-momentum
operators with respect to these angles.

Assuming that the potential $V$ depends only on the magnitude of the
inter-particle separation we write
\begin{equation}
V({\bf r}_1,{\bf r}_2,{\bf r}_3)=V(r_{12}) + V(r_{23}) + V(r_{31}).
\label{eq:Vid}
\end{equation}
We now employ the usual Faddeev decomposition of $\psi$ for three
identical bosons and neglect subsystem angular momentum
\begin{equation}
\Psi(R,\alpha,\Omega)=\psi(R,\alpha_1) + \psi(R,\alpha_2) +
\psi(R,\alpha_3).
\label{eq:decomp}
\end{equation}
Putting (\ref{eq:decomp}) and (\ref{eq:Vid}) into (\ref{eq:hyperschro}) we get
\begin{equation}
(T_R + T_{\alpha_1} - E)\psi(R,\alpha_1) + V(\sqrt{2} R \sin \alpha_1)
  (\psi(R,\alpha_1) + \psi(R,\alpha_2) + \psi(R,\alpha_3))=0.
\end{equation}
Here we have chosen, for definiteness, $k=1$.

We can now exploit the fact that $\psi(R,\alpha_1)$ is independent of
$\hat{r}_{23}$ and $\hat{r}_{1,23}$ to obtain the simplified Faddeev
equation
\begin{equation}
(T_R + T_{\alpha} - E)\psi(R,\alpha)=-V(\sqrt{2} R \sin \alpha)\left(\psi(R,\alpha) + \frac{4}{\sqrt{3}}\int_{|\frac{\pi}{3} -
    \alpha|}^{\frac{\pi}{2} - |\frac{\pi}{6} - \alpha|} \frac{\sin 2
    \alpha'}{\sin 2 \alpha} \psi(R,\alpha') d \alpha' \right).
\label{eq:hyperfad}
\end{equation}
The solution of this equation
can be expanded in a set of eigenfunctions of the hyperangular
operator, i.e.
\begin{equation}
\psi(R,\alpha)=\frac{1}{R^{5/2} \sin(2 \alpha)} \sum_n f_n(R)
\phi_n(R,\alpha),
\end{equation}
where the functions $\phi_n$ satisfy
\begin{equation}
\left[-\frac{\partial^2}{\partial \alpha^2} -
  \lambda_n(R)\right]\phi_n(R,\alpha)=
-\frac{2 M R^2}{\hbar^2} V(\sqrt{2} R \sin
\alpha)\left[\phi_n(R,\alpha)
+
\frac{4}{\sqrt{3}}\int_{|\frac{\pi}{3} -
    \alpha|}^{\frac{\pi}{2} - |\frac{\pi}{6} - \alpha|}
\phi_n(R,\alpha') d \alpha'\right],
\label{eq:phieq}
\end{equation}
with boundary conditions
$\phi_n(R,0)=\phi_n(R,\frac{\pi}{2})=0$. Meanwhile the hyperradial
functions $f_n(R)$ satisfy
\begin{equation}
\left[\frac{\hbar^2}{2M}\left(-\frac{\partial^2}{\partial R^2} + \frac{15}{4
    R^2}\right) + V_n(R)\right]f_n(R)
+ \sum_m \left[2 P_{nm}(R) \frac{\partial}{\partial R} +
  Q_{nm}(R)\right] f_m(R)=Ef_n(R),
\label{eq:hyperrad}
\end{equation}
with the hyperradial potential $V_n(R)$ defined by
\begin{equation}
V_n(R)=\left(\lambda_n(R)-4\right)\frac{\hbar^2}{2 M R^2},
\end{equation}
and $P_{nm}(R)$ and $Q_{nm}(R)$ potentials that induce coupling between
different hyperradial channels~\cite{Braaten:2004rn}:
\begin{eqnarray}
P_{nm}(R)&=&-\frac{\hbar^2}{2M} \sum_k G_{nk}^{-1}(R)
\int_0^{\pi/2} d\alpha \, \phi_k^*(R,\alpha)
\frac{\partial}{\partial R} \phi_m(R,\alpha), \label{eq:Pnm}\\
Q_{nm}(R)&=&-\frac{\hbar^2}{2M} \sum_k G_{nk}^{-1}(R)
\int_0^{\pi/2} d\alpha \, \phi_k^*(R,\alpha)
\frac{\partial^2}{\partial R^2} \phi_m(R,\alpha),
\label{eq:Qnm}
\end{eqnarray}
with 
\begin{equation}
G_{nm}(R)=\int_0^{\pi/2} d\alpha \, \phi_n^*(R,\alpha) \phi_m(R,\alpha).
\end{equation}

Equations~(\ref{eq:Pnm}) and (\ref{eq:Qnm}) make it clear that $P_{nm}$ and
$Q_{nm}$ vanish exactly if the $\phi_n$'s are independent of $R$. As we
shall now show, if the potential $V$ is short ranged, then the
$\phi_n$'s are indeed independent of $R$ \footnote{Below we show that
$P_{nm}(R)$ and $Q_{nm}(R)$ contribute to the energies of the Efimov
spectrum at order $\left(r_s/R \right)^2$.}.

For hyperradii $R$ which are much larger than the range over which $V$
is non-zero, say $l$, we can consider $\alpha$ large enough that
$V(\sqrt{2} R \sin \alpha)$ is zero. The solution of
Eq.~(\ref{eq:phieq}) in this region is then
\begin{equation}
\phi_n^{\rm{(high)}}(\alpha)\approx\sin\left[\sqrt{\lambda_n}\left(\frac{\pi}{2}
    - \alpha\right)\right].
\label{eq:phihigh}
\end{equation}
Note that we have implicitly assumed that $\lambda_n$ becomes
independent of $R$, an assumption which will be justified {\it a
  postieri} in the limit $R \ll |a|$.

On the other hand, as long as we choose $R \gg l$, there is
always a region $\alpha \leq l/(\sqrt{2} R)  \ll 1$. Here
$\hbar^2 \lambda_n(R)/(2 M R^2)$ is much
less than $V(\sqrt{2} R \sin \alpha)$, which typically will be of order
$\hbar^2/(2 M l^2)$.  Consequently for this $\alpha$ domain
the differential equation for $\phi_n(\alpha)$ becomes
\begin{equation}
-\frac{\hbar^2}{2 M R^2} \frac{\partial^2 \phi_n^{\rm (low)}}{\partial \alpha^2}
+ V(\sqrt{2} R \alpha) \phi_n^{\rm (low)}(\alpha)=-\frac{8 \alpha}{\sqrt{3}}
V(\sqrt{2} R \alpha) \phi_n^{\rm (high)}\left(\frac{\pi}{3}\right).
\label{eq:lowalpha}
\end{equation}
Here we have used the fact that $\alpha \ll 1$ in this region to
approximate the integral in Eq.~(\ref{eq:phieq}), and to make the
replacement $\sin \alpha \rightarrow \alpha$. If the derivatives are
all of natural size, then the corrections to this equation are
suppressed by $\alpha^2$. 

The homogenous version of
Eq.~(\ref{eq:lowalpha}) has the solution
\begin{equation}
\phi_n^{\rm{(low})}(\alpha)=A\, \psi_{k=0}(\sqrt{2} R \alpha),
\end{equation}
where $\psi_{k}(r)$ is the wave function of the two-body system for
two-body energy $\frac{\hbar^2 k^2}{2M}$, and $A$x is a constant.
Adding a particular solution of the inhomogeneous
equation we find
\begin{equation}
\phi_n^{\rm{(low})}(\alpha)=A\,\psi_{0}(\sqrt{2} R \alpha)
-\frac{8
  \alpha}{\sqrt{3}} \sin\left(\sqrt{\lambda_n} \frac{\pi}{6} \right)
\label{eq:philow}
\end{equation}
We now consider Eq.~(\ref{eq:lowalpha}) for values of $\alpha$ such
that $l/R \simlt \alpha \ll 1$. In this domain the analysis
that led to Eq.~(\ref{eq:philow})
still applies. But because we are now beyond the range of $V$, we have
\begin{equation}
\psi_k(r)=\frac{\sin(kr + \delta(k))}{k}=\frac{\sin
  \delta(k)}{k}[\cos(kr) + \cot \delta \sin(kr)].
\label{eq:psik}
\end{equation}
As $k \rightarrow 0$ this yields 
$\psi_{0}(r)=r-a$, and we can use this
asymptotic two-body wave function in
Eq.~(\ref{eq:philow}). This gives
\begin{equation}
\phi_n^{\rm (low)}(\alpha)=A (\sqrt{2} R \alpha - a) - \frac{8
  \alpha}{\sqrt{3}}\sin\left(\sqrt{\lambda_n} \frac{\pi}{6}\right).
\end{equation}

But, since $V=0$ in this region, this result
must be consistent with
Eq.~(\ref{eq:phihigh}).  This is achieved by the choice
\begin{equation}
A=-\frac{1}{a} \sin\left[\sqrt{\lambda_n} \frac{\pi}{2}\right],
\label{eq:A}
\end{equation}
which ensures that $\phi_n(\alpha)$ is
continuous across the boundary between ``low'' and ``high'' solutions
at $\alpha \approx l/R$, and the condition
\begin{equation}
\cos\left(\sqrt{\lambda_n} \frac{\pi}{2} \right) -\frac{8}{\sqrt{3 \lambda_n}}
\sin \left(\sqrt{\lambda_n} \frac{\pi}{6}
\right)=\sqrt{\frac{2}{\lambda_n}} \sin\left(\sqrt{\lambda_n}
\frac{\pi}{2}\right) \frac{R}{a},
\label{eq:Danilov}
\end{equation}
on $\lambda_n$, which ensures that $\phi_n(\alpha)$ has a continuous
first derivative as $\alpha \rightarrow l/R$.
We note that if these equations are satisfied $\lambda_n$, and hence
$\phi_n$, is independent of $R$ for $R \ll |a|$, as promised. Indeed, as
long as Eqs.~(\ref{eq:Danilov}) and (\ref{eq:A}) are satisfied the
form (\ref{eq:phihigh}) is the result for $\phi$ for all $\alpha$ such
that $\alpha > l/R$. Solving Eq.~(\ref{eq:Danilov}) in the
limit $R \ll |a|$ we find the lowest eigenvalue
\begin{equation}
\lambda_0=-s_0^2 \left(1 + 1.897 \frac{R}{a} \right),
\end{equation}
with $s_0=1.00624...$. This is the only negative eigenvalue, and hence only this
channel potential is attractive.
So, if we now focus on the unitary limit, where $|a| \rightarrow
\infty$, we have $\lambda_0=-s_0^2$. The hyperradial equation
(\ref{eq:hyperrad}) in this channel then becomes
\begin{equation}
\frac{\hbar^2}{2 M}\left(-\frac{\partial^2}{\partial R^2} - \frac{s_0^2 +
  \frac{1}{4}}{R^2}\right)f_0(R)=Ef_0(R).
\label{eq:hyperradeqLO}
\end{equation}
This equation will hold for $R \gg l$.  If we desire a solution
for negative $E$ the necessity to have $f_0$ be normalizable mandates
that:
\begin{equation}
f^{(0)}_0(R)=\sqrt{R}\, K_{is_0} (\sqrt{2} \kappa R),
\label{eq:LOfn}
\end{equation}
where the superscript $(0)$ indicates that we are working in the
unitary limit, while the subscript $0$ refers to the solution for the
hyperchannel corresponding to $\lambda_0$, which is the only one that
supports bound states. The binding energy of these bound states is
related to the $\kappa$ of Eq.~(\ref{eq:LOfn}) by
\begin{equation}
|E_3| \equiv B \equiv \frac{\hbar^2 \kappa^2}{M}.
\label{eq:Ekappareln}
\end{equation}
Since the attractive $1/R^2$ potential produces a spectrum that is
unbounded from below some other short-distance physics is needed in
order to stabilize the system. If the two-body potential is known this
short-distance physics is provided by the two-body potential $V$, that
becomes operative for $R \sim l$. But an alternative approach is to
add an additional term to Eq.~(\ref{eq:hyperradeqLO}) that summarizes
the impact of the two-body $V$. Here we take this potential to be
a surface delta function at a radius $1/\Lambda$~\cite{Braaten:2004pg}
\begin{equation}
V_{SR}(R)=H_0(\Lambda) \Lambda^2 \delta\left(R - \frac{1}{\Lambda}\right),
\label{eq:VS}
\end{equation}
with $H_0$ adjusted as a function of $\Lambda$ such that the binding
energy of a particular state, say $B_{n*}$ (with a corresponding
$\kappa_*$, given by (\ref{eq:Ekappareln})), is reproduced. Note that
since $V_{SR}$ is operative only at distances $R \sim 1/\Lambda$ it
corresponds to a three-body force. (See Ref.~\cite{Bedaque:1998kg} for
a realization of this in a momentum-space formalism.)

In physical terms we anticipate $l \sim 1/\Lambda$, since we know that
once we consider hyperradii of order $1/\Lambda$ the potential $V$
starts to affect the solutions. Our goal here is to derive {\it
  universal} results, that are independent of details of
$V$. Consequently we will use the procedure described in the previous paragraph
to fix the value of $H_0$ for a given $\Lambda$, and then examine the
residual $\Lambda$-dependence of our results. If we can take the limit
$\Lambda \rightarrow \infty$ without it significantly impacting our
calculations of observables (formally, if all effects scale with
negative powers of $\Lambda$) then these observables are not sensitive
to the physics that takes place in the three-body system for $R \sim
l$. Although we know that our theory is not
valid in the region $R \sim l$, its predictions that have only small ($\sim
1/\Lambda$) corrections due to that short-distance physics will be
``universal'' and not significantly affected by the inaccuracies
in the short-distance region. As we shall see, the
excited-state spectrum is one such observable---which is not
surprising, since these states have very small binding energy
with respect to the scale $l$.

Given that our focus is on predictions of the theory that are
independent of details of $V$ we can consider the extreme case and
take the limit $l \rightarrow 0$. In this limit the form of $K_{is_0}$
as $R \rightarrow 0$ guarantees that once $H_0$ is fixed to give a
bound state at $|E_3|=B_{n*}$, the other binding energies in this
hyperradial eigenchannel form a geometric spectrum. Namely,
$B_n=\hbar^2 \kappa_{n}^2/M$ with
\begin{equation}
\kappa_{n}=\left(e^{-\pi/s_0}\right)^{n-n_*} \kappa_*,
\label{eq:Efimov}
\end{equation}
with $n_*$ the index of the bound state corresponding to $\kappa_*$.
Thus Eq.~(\ref{eq:efispect}) is obtained. 
Eq.~(\ref{eq:Efimov}) will hold for all $\kappa_n$ such that $\kappa_n
\ll \Lambda$. (Note that now the subscript on $\kappa$ denotes the
index of the bound state in adiabatic channel zero.) The continuous
scale invariance of the $1/R^2$ potential has been broken down to a
discrete scale invariance by the imposition of particular short-distance
physics on the problem through the short-distance potential
(\ref{eq:VS})~\cite{Braaten:2004pg}.

\subsection{The Linear Range Correction in the Unitary Limit}

\label{sec-linearcorrn}

The solutions derived in the previous section are valid in the strict
limit $|a| \rightarrow \infty$. In this section we consider the
corrections to this limit that are order $\left(r_s \kappa_*
\right)$. In performing this computation we shall assume that the
range of the two-body potential $l$ is zero~\footnote{One might be
  concerned that this precludes consideration of a positive effective
  range, because of the bound derived in Refs.~\cite{Wi55,PC96}. But
  that bound applies only to energy-independent potentials. If the
  underlying two-body potential $V(R)$ is energy dependent positive
  effective ranges can be generated no matter how small $l$ is.}. The
calculation we are performing therefore corresponds to the limit where
$l \ll |r_s| \ll |a|$. Since the natural scale for the cutoff $\Lambda
\sim 1/l$ we will consider $\Lambda |r_s| \gg 1$.

The first-order correction to EFT results in the $|a| \rightarrow \infty$ limit was already discussed in Ref.~\cite{Hammer:2001gh}. However, in that work Hammer and Mehen considered $r_s/|a|$ effects, and so the first-order corrections they discuss go to zero in the limit $|a| \rightarrow \infty$ that we are considering here. As we shall show here, in principle there is a first-order correction proportional to $r_s \kappa_*$ that could survive if one considers $|a| \rightarrow \infty$ but keeps $r_s$ finite. However, it turns out that the discrete scale invariance that relates wave functions for different bound states at $r_s=0$ guarantees that the particular operator in question has zero matrix element in the limit $\Lambda \rightarrow \infty$. 

To prove this we first observe that in the limit $\Lambda |r_s| \gg 1$ the form (\ref{eq:phihigh}) applies for all values of
$\alpha$, as long as Eq.~(\ref{eq:Danilov}) is obeyed. There
are then two possible sources of corrections to the Efimov spectrum
(\ref{eq:Efimov}) that are linear in $r_s$. First, we must reconsider
Eq.~(\ref{eq:lowalpha}), and take account of the previously neglected
term $\hbar^2 \lambda_n/(2 M R^2)$. As we shall see this leads
to the conclusion that $\phi_n(\alpha)$ is not independent of $R$, and
so then we must also check that $P_{nm}$ and $Q_{nm}$ remain zero at
the level of accuracy we work to here.

Reconsidering Eq.~(\ref{eq:phieq}) we assume that $V \sim
\hbar^2/(2m l^2)$ and consider the region $l/R \simlt \alpha \ll
1$. We can then write
\begin{equation}
\left[-\frac{\partial^2}{\partial \alpha^2} - \lambda_n +\frac{2 M R^2}{\hbar^2}
  V(\sqrt{2} R \sin \alpha)\right] \phi_n^{\rm{(low)}}(\alpha) \approx
  -\frac{2 M R^2}{\hbar^2} V(\sqrt{2} R \sin \alpha)\frac{8
    \alpha}{\sqrt{3}}\phi_n^{\rm{(high)}}\left(R,\frac{\pi}{3}\right).
\label{eq:philownew}
\end{equation}
The approximate form adopted here for the integral on the
right-hand side of Eq.~(\ref{eq:phieq}) is itself accurate to relative
order $\alpha^2$, meaning that the right-hand side omits only terms of
overall order $\alpha^3$. Therefore the first correction to
Eq.~(\ref{eq:philownew}) is $O(l^3/R^3)$ relative to leading. This
equation thus goes one order in $\alpha \sim l/R$ beyond
Eq.~(\ref{eq:lowalpha}). 

In general the homogeneous form of Eq.~(\ref{eq:philownew}) is
difficult to solve. But, for the specific case of $l \rightarrow 0$,
we need only consider the effects of $V$ for $\alpha \rightarrow
0$. There $\sin \alpha=\alpha$ still holds. Hence the solution of the
homogenous form of Eq.~(\ref{eq:philownew}) for any $\alpha > 0$
becomes:
\begin{equation}
\phi_n^{\rm{(low})}(R,\alpha)=A(R) \psi_{k_n}(\sqrt{2} R \alpha),
\label{eq:homog}
\end{equation}
with $\psi_k(r)$ defined by Eq.~(\ref{eq:psik}) and 
$k_n=\frac{\sqrt{\lambda_n}}{\sqrt{2} R}$. Meanwhile, the particular
solution of Eq.~(\ref{eq:philownew}) is unchanged up to the accuracy
to which  we work here, and so
\begin{equation}
\phi_n^{\rm(low)}(R,\alpha)=A(R) \psi_{k_n}(\sqrt{2} R \alpha) - \frac{8 \alpha}{\sqrt{3}}
\sin\left(\sqrt{\lambda_n} \frac{\pi}{6}\right) + O(\alpha^3).
\label{eq:inhomog}
\end{equation}
Substituting in the form (\ref{eq:psik}) for $\psi_{k_n}(r)$ we see that
$\phi_n^{\rm(low)}$ will still match smoothly with 
the large-$\alpha$ solution (\ref{eq:phihigh}), provided that
\begin{equation}
\frac{A(R)}{k_n} \sin\left(\delta\left(k_n\right)\right)
=\sin\left(\sqrt{\lambda_n} \frac{\pi}{2}\right),
\end{equation}
and
\begin{eqnarray}
\cos\left(\sqrt{\lambda_n} \frac{\pi}{2} \right)
-\frac{8}{\sqrt{3 \lambda_n}} 
\sin \left(\sqrt{\lambda_n} \frac{\pi}{6} \right)&=&-R \, \, k_n \cot(\delta(k_n)) \, 
\sqrt{\frac{2}{\lambda_n}} \sin \left(\sqrt{\lambda_n}
  \frac{\pi}{2}\right) \\
&=&\left[\frac{R}{a} - \frac{\lambda_n}{4} \frac{r_s}{R} + O\left(\frac{l^3}{R^3}\right)\right]\sqrt{\frac{2}{\lambda_n}} \sin \left(\sqrt{\lambda_n}
  \frac{\pi}{2}\right).
\label{eq:Danilov-rs}
\end{eqnarray}
In Eq.~(\ref{eq:Danilov-rs}) we have assumed only that the coefficients of all terms in the effective-range expansion after the $O(k^2)$ one scale with $l$.
Under this assumption the scattering length and effective range determine the eigenvalue $\lambda_n$, up to corrections suppressed by $(l/R)^3$.

 In this case Eq.~(\ref{eq:Danilov-rs}) can
be solved by noting that $ \lambda_n(R)$ has a self-consistent solution
in the region $|r_s| \ll R \ll |a|$
\begin{equation}
\lambda_n(R)=\lambda_n^{(0)}(0) \left(1 + \gamma_n \frac{R}{a}  +\xi_n \frac{r_s}{R} + \ldots \right),
\label{eq:lambda_n-rs}
\end{equation}
with $\lambda_n^{(0)}(R)$ is the solution of Eq.~(\ref{eq:Danilov}), which here we need only at $R=0$.
By a perturbative evaluation of
Eq.~(\ref{eq:Danilov-rs}) in powers of $r_s$ we find that:
\begin{equation}
\xi_n = - \frac{\lambda_n^{(0)}(0) \gamma_n}{4} 
\label{eq:xi_n}
\end{equation}
encodes the $O(r_s)$ correction to $\lambda_n$ in the limit $|a|
\rightarrow \infty$. 
Hence, the value of the lowest eigenvalue is
\begin{equation}
\lambda_0(R)= -s_0^2 \left(1 + 1.897 \frac{R}{a} + 0.480 \frac{r_s}{R} + \ldots \right),
\label{eq:lambda0rs}
\end{equation}
where we have omitted corrections of second order in $R/a$ and
$r_s/R$ and have not considered $r_s/a$ corrections. 
Similarly, the second-lowest eigenvalue can be determined as:
\begin{equation}
\lambda_1 (R) = 19.94 \left(1 - 0.1872 \frac{R}{a} + 0.9333 \frac{r_s}{R} + \ldots \right).
\label{eq:lambda1rs}
\end{equation}
In what follows, we will focus only on the lowest eigenvalue $\lambda_0(R)$.

With $\lambda_0(R)$ in hand we can compute
\begin{equation}
\phi_0(\alpha;R)=\sin\left[\sqrt{\lambda_0(R)}\left(\frac{\pi}{2}
    - \alpha\right)\right].
\end{equation}
for $\alpha > l/|a|$, up to corrections of order
$\left(r_s/R\right)^2$, with $\lambda_0$ given by
Eq.~(\ref{eq:lambda0rs}). For later purposes it makes sense to
normalize $\phi_0(\alpha;R)$ such that~\cite{Macekold}
\begin{equation}
\int_0^{\frac{\pi}{2}} \phi_n^2(\alpha;R) \, d \alpha=1,
\end{equation}
irrespective of the value of $R$.
Since the overall normalization of $\phi_n$ is arbitrary this is easily done,
resulting in:
\begin{equation}
\phi_0(\alpha;R)=\frac{1}{N(R)} \sin\left[\sqrt{\lambda_0(R)}\left(\frac{\pi}{2}
    - \alpha\right)\right],
\label{eq:phi0norm}
\end{equation}
with
\begin{equation}
N^2(R)=\frac{\pi}{4} \left[1-j_0(\pi \sqrt{\lambda_0(R)})\right]
\end{equation}
where $j_0$ is the spherical Bessel function of zeroth order.

The fact that $\lambda_n$ now depends on $R$ opens the possibility
that the hyperchannel coupling potentials $P_{nm}$ and $Q_{nm}$ will
no longer be zero. Clearly both can be at most
$O\left(\frac{r_s}{R}\right)$, and so at leading order in $r_s$ we
need only consider $P_{nn}(R)$ and $Q_{nn}(R)$. Given that our focus
here is on the bound-state spectrum this means we only have to examine $P_{00}(R)$ and $Q_{00}(R)$. 
The use of normalized $\phi_0(\alpha;R)$
can then easily be seen to
guarantee that $P_{00}(R)$ is identically zero~\cite{Macekold}.  A brief computation also shows that
\begin{equation}
Q_{00}(R)=0+O\left(\frac{r_s^2}{R^2}\right),
\end{equation}
and so we need not concern ourselves further with these adiabatic
coupling potentials in pursuing our calculation of effects linear in $r_s$.

Consequently, in the unitary ($|a| \rightarrow \infty$), but not scaling ($r_s \neq
0$) limit the hyperradial Eq.(\ref{eq:hyperrad}) becomes
\begin{equation}
\frac{\hbar^2}{2 M}\left(-\frac{\partial^2}{\partial R^2} - \frac{s_0^2 +
  \frac{1}{4}}{R^2} -\frac{s_0^2\, \xi_0 \,r_s}{R^3}
\right)f^{(1)}(R)=E^{(1)} f^{(1)}(R)~,
\label{eq:hyperradialfirstorder}
\end{equation}
where the superscript one indicates that we have worked only to first
order in $r_s$ . The form of
Eq.~(\ref{eq:hyperradialfirstorder}), but not the specific pre-factor
of the $1/R^3$ potential, was proposed by Efimov on dimensional
grounds~\cite{Efimov91}. 

The momentum-space evaluation of first-order (in $r_s/|a|$) effects in Ref.~\cite{Hammer:2001gh} neglected the operator corresponding to the $1/R^3$ potential that appears in Eq.~(\ref{eq:hyperradialfirstorder}). The authors of Ref.~\cite{Hammer:2001gh} argued that the piece of the two-body scattering amplitude that survives in the limit $|a| \rightarrow \infty$ and is proportional to $r_s$ does not have a two-body bound-state pole and so cannot lead to effects in the three-body system. As we shall now see the conclusion that the $1/R^3$ potential does not affect the spectrum in first-order perturbation theory in the unitary limit is correct, although it is {\it not} justified to neglect this piece of the three-body dynamics entirely, as was done in Ref.~\cite{Hammer:2001gh}. 

Now---to the extent that the $r_s/R^3$ term is a perturbation---the
shift in the Efimov spectrum near a Feshbach resonance (i.e. near $|a|=\infty$)
can be evaluated by computing the matrix elements of this
perturbing $1/R^3$ potential between leading-order hyperradial
functions (\ref{eq:LOfn}).  In other words, we have, for the
first-order shift in the bound-state energy:
\begin{equation}
\frac{2 M}{\hbar^2} \,\Delta B^{(1)}_{n}=s_0^2 \,r_s \,\xi_0 \int dR
{f_n^{(0)}}^2(R) \frac{1}{R^3},
\label{eq:shifta}
\end{equation}
with $E_n^{(1)}=-(\frac{\hbar^2 \kappa_n^2}{M} + \Delta B^{(1)}_n)$ the
energy of the $n$th bound state up to first order in $r_s$. The
$f_n^{(0)}$ that appears in Eq.~(\ref{eq:shifta}) is the normalized
zeroth-order hyperradial wave function of the corresponding bound
state~\cite{Braaten:2004rn}:
\begin{equation}
f^{(0)}_n(R)=2 \kappa_n \sqrt{\frac{\sinh(\pi s_0)}{\pi s_0}}
R^{1/2}  K_{is_0}(\sqrt{2} \kappa_n R).
\label{eq:f0n}
\end{equation}

Unfortunately the integral in Eq.~(\ref{eq:shifta}) is divergent. This divergence can,
however, be absorbed by a modification of the short-distance potential
(\ref{eq:VS}) to include terms $\sim r_s$. I.e. we now add to
$V_{SR}(R)$ of Eq.~(\ref{eq:VS})
a piece:
\begin{equation}
V^{(1)}_{SR}(R)=H_1(\Lambda) \Lambda^2 \delta\left(R-\frac{1}{\Lambda}\right),
\end{equation}
with $H_1 \sim r_s$.

We include this potential in Eq.~(\ref{eq:hyperradialfirstorder}), and
then treat it in perturbation theory. Perturbation theory is only manifestly
valid if $|r_s| \Lambda \ll 1$. 
But, as we will show below, after renormalization we can take the limit $\Lambda \rightarrow \infty$ in our perturbation-theory
  expressions. Although the
  $1/R^3$ potential and $V_{SR}^{(1)}$ cannot individually be treated in perturbation theory 
  in this limit their combined effects remain perturbative for arbitrarily large $\Lambda$.
  Indeed, we will show that the $\Lambda$
dependence in the perturbation-theory calculation is confined to
terms of $O(1/\Lambda)$ and so it is always small. The complete
first-order-perturbation-theory result is therefore independent of details of
short-distance dynamics, i.e. it is ``universal'' in the sense we
are using the term here.

At finite $\Lambda$ the total first-order shift in the energy of the
$n$th bound state in the zeroth eigenchannel is then:
\begin{equation}
\frac{2 M}{\hbar^2} \Delta B^{(1)}_n=s_0^2 r_s \xi_0 \left[\int_{\frac{1}{\Lambda}}^\infty dR
{f^{(0)}_n}^2(R) \frac{1}{R^3} - \frac{h_1}{2} \Lambda^2
{f^{(0)}_n}^2\left(\frac{1}{\Lambda}\right)\right],
\label{eq:foshifteq1}
\end{equation}
where the theory has been regulated at a distance $1/\Lambda$ and we
shall seek to remove this cutoff (i.e. take $\Lambda \rightarrow
\infty$) at the end of the calculation. We shall, however, calculate
throughout in the limit that $\Lambda \gg \kappa_*$ (indeed, that
$\Lambda$ is much larger than the binding momentum of any states we
are interested in). The dimensionless $h_1$ in
Eq.~(\ref{eq:foshifteq1}) is defined as:
\begin{equation}
h_1=\frac{4 H_1 M}{\hbar^2 s_0^2 \,r_s \,\xi_0},
\end{equation}
and the subscripts on the hyperradial wave functions indicate which
binding energy $B_n$ they correspond to. From now on we will drop the
superscript on these $f_n$'s, since for the rest of this Subsection
all wave functions are to be evaluated at leading order.

Substituting for $f_n$ from Eq.~(\ref{eq:f0n}), and using the
short-distance form of $K_{is_0}(x)$:
\begin{equation}
K_{is_0}(x)=-\sqrt{\frac{\pi}{s_0 \sinh(\pi s_0)}}\sin
\left(s_0 \ln \left(\frac{x}{\sqrt{2}}\right) + \alpha_0\right) + O(x^2),
\label{eq:Kis0smallx}
\end{equation}
with~\cite{Braaten:2004rn}
\begin{equation}
\alpha_0 = -\frac{1}{2} s_0 \ln 2 - \frac{1}{2} \arg \left(\frac{\Gamma(1 +
  is_0)}{\Gamma(1- is_0)}\right),
\end{equation}
we get:
\begin{equation}
\frac{2 M \Delta B^{(1)}_n}{\hbar^2}=4 r_s \xi_0 \kappa_n^2 \left[ \frac{s_0 \sinh (\pi  s_0)}{\pi} \int_{\frac{1}{\Lambda}}^\infty dR
\frac{K_{is_0}^2(\sqrt{2} \kappa_n R)}{R^2} - \frac{h_1(\Lambda) \Lambda}{2} \sin^2\left(s_0 \ln
\left(\frac{\kappa_n}{\Lambda}\right) + \alpha_0\right) \right].
\label{eq:divergent}
\end{equation}

The result of the integral inside the square brackets can be expressed in terms of Hypergeometric functions. Since we are only interested in the result for $\Lambda \gg \kappa_n$ we expand these functions in powers of $\kappa_n/\Lambda$ and keep the first two terms to obtain:
\begin{equation}
I(z) \equiv \int_z^{\infty} \, \frac{dx}{x^2} K_{is_0}^2(x)=\frac{1}{z} I_{-1}(z) + z I_1(z) + O(z^3),
\label{eq:Iexp}
\end{equation}
with the functions $I_{-1}$ and $I_1$ encoding non-analytic dependence of the coefficients on $z$:
\begin{eqnarray}
I_{-1}(z)&=& \frac{\pi}{2 s_0 \sinh(\pi s_0)}\left[1- \frac{1}{\sqrt{1 + 4 s_0^2}}\cos\left(2 s_0 \ln(z/\sqrt{2}) + 2 \alpha_0 + \arctan(2 s_0)\right)\right],\label{eq:Im1}\\
I_1(z)&=&-\frac{\pi}{4 s_0 (1+s_0^2) \sinh(\pi s_0)} \left[1 - \sqrt{\frac{1+s_0^2}{1 + 4 s_0^2}} \right.\nonumber\\
&& \qquad \qquad \qquad \qquad \times \left.\cos\left(2 s_0 \ln(z/\sqrt{2}) + 2 \alpha_0 - \arctan\left(\frac{3s_0}{1-2s_0^2}\right)\right)\right].
\end{eqnarray}
Inserting Eq.~(\ref{eq:Iexp}) into the expression for the energy shift we have:
\begin{eqnarray}
\frac{2 M \Delta B^{(1)}_n}{\hbar^2}=4 r_s \xi_0
  \kappa_n^2 \left[\frac{s_0 \sinh(\pi s_0)}{\pi} \left(\Lambda I_{-1}\left(\frac{\sqrt{2} \kappa_n}{\Lambda}\right)+ \frac{2 \kappa_n^2}{\Lambda} I_1\left(\frac{\sqrt{2} \kappa_n}{\Lambda}\right) + \ldots\right)\right.\nonumber\\
\left. - \frac{h_1(\Lambda) \Lambda}{2} \sin^2\left(s_0 \ln
\left(\frac{\kappa_n}{\Lambda}\right) + \alpha_0\right) \right]
 \label{eq:DeltaBnrenorm}
\end{eqnarray}

It follows from Eqs.~(\ref{eq:DeltaBnrenorm}) and (\ref{eq:Im1}) that the coefficient of the linear divergence can be forced to zero if we choose:
\begin{eqnarray}
h_1(\Lambda)= \frac{1}{\sin^2(s_0 \ln x_0 + \alpha_0)}
\left[1-\frac{1}{\sqrt{1 + 4 s_0^2}}\cos\left(2 s_0 \ln(x_0) + 2 \alpha_0 + \arctan(2 s_0)\right)\right]{\label{eq:h1}}
\end{eqnarray}
with $x_0 \equiv {\kappa_*}/{\Lambda}$. 
With this choice of $h_1$ we can write:
\begin{equation}
\frac{2 M \Delta B^{(1)}_n}{\hbar^2}=4 r_s \xi_0 \kappa_n^2
\left [\Lambda f\left(\frac{\kappa_n}{\Lambda}\right) + \frac{\kappa_n^2}{\Lambda} g\left(\frac{\kappa_n}{\Lambda}\right)\right],
\label{eq:DeltaBnrenorm2}
\end{equation}
where the function $g$ is proportional to $I_1(\sqrt{2} \kappa_n/\Lambda)$ up to small corrections and so is generically of order one---although it does have log-periodic dependence on $\kappa_n/\Lambda$. Similarly the function $f$
inherits the log periodicity of $I_{-1}$:
\begin{equation}
f(z e^{\frac{\pi}{s_0}})=f(z),
\label{eq:logperiodic1}
\end{equation}
while the renormalization condition guarantees:
\begin{equation}
f(x_0)=0.
\label{eq:frenorm}
\end{equation}
Since, at leading order, $\kappa_n=\kappa_* e^{\frac{(n_* - n)  \pi}{s_0}}$, Eqs.~(\ref{eq:logperiodic1}) and (\ref{eq:frenorm}) result in:
\begin{equation}
f\left(\frac{\kappa_n}{\Lambda}\right)=f\left(\frac{\kappa_*}{\Lambda}\right)=0.
\label{eq:logperiodic}
\end{equation}
Therefore once the linearly divergent part of the shift is renormalized away for one state, it will---at first order in perturbation theory---be zero for all bound states in the Efimov spectrum.
Consequently,
\begin{equation}
\Delta B_n^{(1)}=\frac{4 \xi_0 r_s \kappa_n^4}{\Lambda} g\left(\frac{\kappa_n}{\Lambda}\right).
\label{eq:DeltaBnfinal}
\end{equation}
Numerical results generated for the first-order shift to the spectrum in the unitary limit conform to this pattern~\cite{Daekyoung}. Eq.~(\ref{eq:DeltaBnfinal}) holds as long as we renormalize in such a way as to keep $\kappa_*$ fixed. 

Therefore a renormalization procedure that leaves one of the Efimov states unperturbed by $r_s$ corrections leaves all states
unperturbed in the limit $\Lambda \rightarrow \infty$.
This conclusion is a consequence of the discrete
scale invariance of the Efimov spectrum. Because the leading-order
spectrum has discrete scale invariance, and states are related to
one another by a scale transformation of a factor
$e^{(n-n_*)\frac{\pi}{s_0}}$, the integrals (\ref{eq:shifta}) for
different states $n$ are also related to one another by that factor,
i.e.
\begin{equation}
\Delta B^{(1)}_{n}=e^{\frac{-3 \pi (n-n_*)}{s_0}} \Delta B^{(1)}_{n_*}.
\end{equation}
Thus, if $\Delta B^{(1)}_{n_*}$ is forced---by our renormalization
procedure---to equal zero, we must also have
\begin{equation}
\Delta B^{(1)}_{n}=0~\mbox{for all $n$.}
\label{eq:noshift}
\end{equation}
In order for this conclusion to hold it is critical to use a
regularization and renormalization prescription which respects
discrete scale invariance. The cutoff-plus-delta-function procedure we
adopted does this in the limit $\Lambda \rightarrow \infty$. At any
finite $\Lambda$ the violations of discrete scale invariance result
in $1/\Lambda$-suppressed corrections to Eq.~(\ref{eq:noshift}). However, even if $\Lambda$ is kept $\sim 1/l$, the perturbative corrections to the spectrum would only be $\sim r_s l \kappa_n^4$, and so is of higher order than we are considering here.

\section{The Linear Range Correction for arbitrary Scattering Length}
\label{sec-pspace}

In the previous section we focused on the limit in which $|a| \rightarrow
\infty$, i.e. $\kappa_* \gg 1/|a|$. The hyperradial
formalism is particularly well-suited to the analysis of that limit, as the hyperradial potential is a power law (or, more generally, sum of powers of $r_s/R$) for all values of $R$. In this
section we use a momentum-space formalism~\cite{Afnan:2003bs} to obtain the corrections to the Efimov spectrum for arbitrary values of
$1/(a \kappa_*)$.
Numerical calculation of scattering and bound-state observables is
straightforward, and it is easy to compute range corrections---regardless of the value of
$a$~\cite{Platter:2006a}. We review the relevant formalism in Section~\ref{sec-smserev}. Results for the range correction are presented in Sec.~\ref{sec-ren_kappa} and analyzed for their content in terms of universality in Sec.~\ref{sec-smseuniv}. 

A technical point is that to do this we employ the effective-range expansion around the bound state pole
\beq
k\cot \delta_0 =-\gamma+\frac{\rho}{2}(k^2+\gamma^2)+\ldots~,
\label{eq:erebspole}
\eeq
rather than the one around $k=0$, since Eq.~(\ref{eq:erebspole})
facilitates the treatment of the two-body
threshold cut generated by the three-body scattering equations. The
parameter $\rho$ is equal to the effective range $r_s$ up to corrections $\sim r_s^2/a$, 
which means that setting $\rho=r_s$ is adequate for our NLO calculation. Therefore in what
follows we do not distinguish between $\rho$ and $r_s$.

\subsection{Subtracted Momentum Space Equations: A Review}
\label{sec-smserev}
For zero-range, $l=0$, two-body potentials the usual momentum-space
Faddeev equations for the atom-dimer scattering amplitude reduce to
the equation first derived by Skorniakov and Ter-Martirosian (STM)~\cite{STM57}~\footnote{The STM form appears different, but is equivalent~\cite{Bedaque:1998kg,Afnan:2003bs}.}:
\beq
\label{eq:stm}
\mathcal{A}^{(0)}(q,q';E)=\mathcal{Z}(q,q';E)+\int_0^\Lambda\hbox{d}q''\,q''^2
\mathcal{Z}(q,q'',E)\frac{S^{(0)}(E;q'')}{E+E_D-\frac{3}{4}\frac{\hbar^2 q''^2}{M}+i\varepsilon}
\mathcal{A}^{(0)}(q'',q';E)~,
\eeq where $\mathcal{A}^{(0)}$ denotes the leading-order atom-dimer
scattering amplitude, $E_D$ is the
dimer binding energy, $\Lambda$ a
momentum cutoff which regularizes the integral equation, $\varepsilon$ a positive infinitesimal, and
\beq
\mathcal{Z}(q,q';E)=-\frac{M}{q q' \hbar^2} \log\left(\frac{q^2+q'^2+q
  q'-\frac{ME}{\hbar^2}}{q^2+q'^2-q q'-\frac{ME}{\hbar^2}}\right).
\eeq
Finally, in Eq.~(\ref{eq:stm}), $S^{(0)}(E;q'') \equiv S^{(0)}\left(E-\frac{3 \hbar^2 q''^2}{4m}\right)$ denotes the function:
\beq
S^{(0)}(E)=\frac{2 \hbar^4}{\pi M^2} (\gamma - i k)
\label{eq:tauk}
\eeq with $E_D=\hbar^2 \gamma^2/M$ and $E=\hbar^2 k^2/M$. Note that here we have
$E_D$ (equivalently $|a|$) finite, in contrast to the analysis of the
previous section. Note also that if $\gamma < 0$ then the pole
corresponding to the dimer is on the second sheet of the complex
energy plane, and so does not correspond to an actual bound state, but
instead to a ``quasi-bound state''. Regardless of the sign of $\gamma$
Eq.~(\ref{eq:stm}) may be made more compact by introducing the 
dimer propagator
\beq 
\tau^{(0)}(E) \equiv
\frac{S^{(0)}(E)}{E+E_D + i \epsilon}.
\label{eq:LOtau}
\eeq

Although now 50 years old the STM equation has recently been
re-derived as the leading-order result for the atom-dimer scattering
amplitude in an effective field theory (EFT) with contact interactions
alone. This EFT is the appropriate low-energy theory for
non-relativistic systems with $a \gg l$ and $\kappa\, l \ll 1$, where
$\kappa$ denotes the momentum scale of observable of interest \cite{Bedaque:1998kg}.

The EFT analysis of Ref.~\cite{Bedaque:1998kg} showed that
Eq.~(\ref{eq:stm}) leads to strongly cutoff-dependent results for the
bound-state spectrum. In coordinate space this is reflected by the
necessity to fix the phase of the hyperradial wave function at short
distances before a prediction about the three-body bound-state
spectrum can be made. As in the analysis of Sec.~\ref{sec-rspace}, it
transpires that fixing a single three-body observable (in what follows
we choose the threshold atom-dimer scattering amplitude) is sufficient
to remove the cutoff dependence of the spectrum. We will now
show how to achieve this using a once-subtracted version of the
integral equation (\ref{eq:stm}). Our presentation follows Ref.~\cite{Afnan:2003bs}.

We begin by considering:
\beq
\mathcal{A}^{(0)}(q,0;-E_D)=\mathcal{Z}(q,0;-E_D)-\frac{4 M}{3}\int_0^\Lambda\hbox{d}
q''\,\mathcal{Z}(q,q'';-E_D)S^{(0)}(-E_D,;q'')\mathcal{A}^{(0)}(q'',0;-E_D)~,
\label{eq:threshoffshell}
\eeq
We can insert the required three-body input by demanding
that the scattering amplitude reproduces the correct particle-dimer
scattering length at threshold. The on-shell amplitude at $E=-E_D$
obeys Eq.~(\ref{eq:threshoffshell}) with $q=0$, but the value of
$\mathcal{A}^{(0)}(q,0;-E_D)$ is fixed if the atom-dimer scattering
length $a_3$ is taken as input. Therefore by subtracting the equation
at $q=0$ from Eq.~(\ref{eq:threshoffshell}) and
relating $\mathcal{A}^{(0)}$ to $a_3$
we obtain
the subtracted equation at threshold:
\bea
\mathcal{A}^{(0)}(q,0;-E_D)&=&\frac{3 M a_3}{8 \gamma}
+\Delta[\mathcal{Z}](q,0;-E_D)
\nonumber\\[0.3cm]
&&\quad-\frac{4 M}{3}\int_0^\Lambda\hbox{d}q''
\Delta[\mathcal{Z}](q,q'';-E_D)
S^{(0)}(-E_D;q'')\mathcal{A}^{(0)}(q'',0;-E_D).
\label{eq:threshsubt}
\eea Here \beq
\Delta[\mathcal{Z}](q,q';E)=\mathcal{Z}(q,q';E)-\mathcal{Z}(0,q';E),
\eeq
and $a_3$ denotes the atom-dimer scattering length. 

Now we can determine the full-off-shell amplitude at threshold by exploiting
the symmetries of $\mathcal{A}$. Before subtraction
the full off-shell amplitude at threshold satisfies
\beq
\label{eq:full-off-shell-not-renormalized}
\mathcal{A}^{(0)}(q,q';-E_D)=\mathcal{Z}(q,q';-E_D)
-\frac{4 M}{3}\int_0^\Lambda\hbox{d}q''\,\mathcal{Z}(q,q'';-E_D)
S^{(0)}(-E_D;q'')\mathcal{A}^{(0)}(q'',q';-E_D).
\eeq
Since
$\mathcal{Z}(q,q';E)=\mathcal{Z}(q',q;E)$, we also have
$\mathcal{A}^{(0)}(0,q;-E_D)=\mathcal{A}^{(0)}(q,0;-E_D)$. 
Using this fact as input, together with the solution of
Eq.~(\ref{eq:threshsubt}), produces the subtracted version of
Eq.~(\ref{eq:full-off-shell-not-renormalized}):
\bea
\nonumber
\mathcal{A}^{(0)}(q,q';-E_D)&=&{\cal A}^{(0)}(0,q';-E_D)+\Delta[\mathcal{Z}](q,q';-E_D)\\[0.3cm]
&\quad&-
\frac{4 M}{3}\int_0^\Lambda\hbox{d}q'' 
\Delta [\mathcal{Z}](q,q'';-E_D)S^{(0)}(-E_D;q'')\mathcal{A}^{(0)}(q'',q';-E_D),
\label{eq:fulloffshell}
\eea
which determines the fully-off-shell amplitude at threshold.

With Eq.~(\ref{eq:fulloffshell}) in hand resolvent identities may be
used to obtain the subtracted amplitude at any energy~\cite{Afnan:2003bs}:
\beq
\label{eq:full-amplitude-subtracted}
\mathcal{A}^{(0)}(q,k;E)=\mathcal{A}^{(0)}(q,k;-E_D)+B^{(0)}(q,k;-E_D)+
\int_0^\Lambda\hbox{d}q' q'^2\,Y^{(0)}(q,q';E) \mathcal{A}^{(0)}(q',k;E)~,
\eeq
where the second inhomogeneous term is given by
\beq
B^{(0)}(q,k;E)=\delta[\mathcal{Z}](q,k;E)+\int_0^\Lambda\hbox{d}q''\,{q''}^2\,
\mathcal{A}^{(0)}(q,q'';-E_D) \, \tau^{(0)}(-E_D;q'')\delta[\mathcal{Z}](q'',k;E)~,
\eeq
with
\beq
\delta[\mathcal{Z}](q,q';E)=\mathcal{Z}(q,q';E)-\mathcal{Z}(q,q';-E_D)~,
\eeq
and $\tau(E;q)=\tau(E-{\textstyle\frac{3\hbar^2 q^2}{4 M}})$. The kernel in
Eq.(\ref{eq:full-amplitude-subtracted}) is
\bea
\nonumber
Y^{(0)}(q,q';E)&=&\mathcal{A}^{(0)}(q,q';-E_D)\delta[\tau^{(0)}](E;q')
+\delta[\mathcal{Z}](q,q';E)
\tau^{(0)}(E;q')\\
\nonumber
&&\qquad+\int_0^\Lambda\hbox{d}q'' \,q''^2\,\mathcal{A}^{(0)}(q,q'';-E_D)
\tau^{(0)}(-E_D;q'')
\delta[\mathcal{Z}](q'',q';E)\tau^{(0)}(E;q')\\
&=&\mathcal{A}^{(0)}(q,q';-E_D)\delta[\tau^{(0)}](E;q')+B^{(0)}(q,q';E)\tau^{(0)}(E;q')~,
\label{eq:Y0}
\eea
with:
\beq
\delta[\tau^{(0)}](E;q')=\tau^{(0)}(E;q') - \tau^{(0)}(0;q').
\label{eq:deltatau}
\eeq

The result is a formulation of the three-body problem with short-range
forces in which only renormalized quantities appear. Subtracted
equations relate the amplitude at different energies (including
energies below the atom-dimer scattering threshold).  Hence, once
$a_3$ is known, the three-body bound-state spectrum and atom-dimer
phase shifts can be computed.

For pure S-wave interactions the only impact of effective-range
corrections on the above argument is a modification of the two-body
propagator $\tau$~\cite{Platter:2006a}. For arbitrary $r_s$ this can
be written in the form
\beq \tau(E)=-\frac{2 \hbar^2}{\pi
  M^2}\frac{1}{-\gamma - i k + \frac{r_s}{2}(\gamma^2+k^2)}~. 
\eeq
This two-body propagator has one additional
pole which lies outside the region of validity of the EFT. Therefore,
the two-body propagator has to be expanded up to the desired order in
$r_s/a$~\footnote{Technically, $r_s$ must be replaced by $\rho$ if we wish
to consider any order beyond $n=1$.}:
\beq
\tau^{(n)}(E)=\frac{S^{(n)}(E)}{E+E_D+i\varepsilon}~,
\eeq
with $S^{(n)}$ for $n<3$
\beq S^{(n)}(E)=\frac{2 \hbar^4}{\pi
  M^2}\sum_{i=0}^n\left(\frac{r_s}{2}\right)^i[\gamma - ik]^{i+1}~.
\label{eq:Sn}
\eeq
This form for $S^{(n)}$ is then inserted into the above equations and
the calculation at NLO and N2LO can proceed exactly as with the LO
calculation. Below we denote the resulting amplitude by ${\cal A}^{(n)}$.

\subsection{Renormalization to $\kappa_*$}
\label{sec-ren_kappa}
When the subtracted integral equations are presented as has been done in Subsection~\ref{sec-smserev}, it appears that to use them we must choose $a_3$ as the observable that provides the three-body input to the scattering equation. This
constitutes a renormalization scheme that is appropriate for the
low-energy properties of a system with a fixed atom-atom scattering
length.  In the situations of interest to us here though, the two-body
scattering length of the atomic species is controlled by an external
magnetic field, through the phenomenon of Feshbach resonances.  The
long-range properties of the atomic system (including the three-body
scattering length) can then vary rapidly if the resonance is
narrow. But, physically, the
short-distance properties of the three-body system---in particular
the asymptotic form of the three-body wave function in the
high-momentum/short-distance region---should be unaffected by
changes in the infra-red physics associated with the two-body
scattering length. In what follows we will therefore use a
renormalization scheme in which the parameter associated with this ultra-violet
part of the wave function is independent of the two-body scattering
length. The analysis of Sec.~\ref{sec-rspace} shows that (in the
unitary limit) the phase of this part of the wave function is in
one-to-one correspondence with the binding momentum of the $n*$th three-body bound
state, $\kappa_*$, and so this renormalization scheme amounts to
ensuring that the value of $\kappa_*$, taken at the point
$\gamma=0$, is unchanged when we include corrections that are proportional to
$r_s$.

This means we need to know how to relate observables calculated in the
unitary limit, $\gamma=0$, to observables calculated at a finite value of $\gamma$.
More generally, we would like to develop renormalized equations that
encode relations between the amplitude ${\cal A}^{(n)}$ at different
values of the two-body binding momentum $\gamma$. The
$\gamma$-dependence of ${\cal A}^{(n)}$ arises in the integral equation
through the dependence of $\tau$ and ${\cal Z}$ on the two-body
scattering length~\footnote{${\cal Z}$ is $\gamma$-dependent at fixed
on-shell momentum $k$, which is how we formulate our scattering problem.}. In
what follows we use this fact, together with the subtraction approach,
to relate the atom-dimer scattering amplitudes at two different 
values of $\gamma$. Choosing
$\gamma=0$ as one point in the resulting equation then allows us to
compute ${\cal A}^{(n)}$ for any desired value of $\kappa_*$.

Consider the operator form of the integral equation for the
threshold scattering-amplitude ${\cal A}$  \beq
\label{eq:stmgam}
\mathcal{A}(\gamma,0)=\mathcal{Z}(\gamma,0)+
\mathcal{Z}(\gamma,0)\,\tau(\gamma,0)\,\mathcal{A}(\gamma,0)~, 
\eeq
where now we have written the $\gamma$-dependence explicitly, as the
first argument in all quantities, and indicated that this is an
equation at threshold by including the atom-dimer relative momentum,
$k$ as a second argument. This supersedes the $k=\sqrt{ME}/\hbar$ employed above, and the argument $E$ used there is now
related to $k$ by 
\begin{equation}
E=\frac{3 \hbar^2 k^2}{4M} - E_D.
\end{equation}
At the atom-dimer scattering
 threshold $k=0$. We have also suppressed the superscript $n$ in Eq.~(\ref{eq:stmgam}) since the argument we develop in the next few paragraphs
is independent of the $\tau^{(n)}$ that is employed there. 

Equation (\ref{eq:stmgam}) can be written as \beq
\label{eq:subtraction_efimov_1}
\mathcal{Z}(\gamma,0)=\mathcal{A}(\gamma,0)[1+\tau(\gamma,0)\mathcal{A}(\gamma,0)]^{-1}~.
\eeq Assuming that we have already calculated the scattering amplitude
for a given combination of $a_3$ and a certain $\gamma$, say $\gamma_1$, using
Eq.~(\ref{eq:threshsubt}) and (\ref{eq:fulloffshell}), we can now use
Eq.(\ref{eq:subtraction_efimov_1}) to derive the scattering amplitude
for the two-body scattering length that corresponds to
$\gamma_2$. Evaluating the expression for
$\mathcal{Z}(\gamma_1,0)-\mathcal{Z}(\gamma_2,0)$ gives 
\bea \nonumber
\mathcal{Z}(\gamma_1,0)-\mathcal{Z}(\gamma_2,0)&=&
        [1+\mathcal{A}(\gamma_1,0)\tau(\gamma_1,0)]^{-1}{\cal
          A}(\gamma_1,0)\\
        &&\qquad\qquad-\mathcal{A}(\gamma_2,0)[1+\tau(\gamma_2,0)
          \mathcal{A}(\gamma_2,0)]^{-1}~.  \eea Now multiplying
        with the appropriate expression from the left- and right-hand
        side, we obtain 
an integral equation for
        $\mathcal{A}(\gamma_2,0)$ 
\bea \nonumber
        \mathcal{A}(\gamma_2,0)&=&{\cal A}(\gamma_1,0)
        +\mathcal{A}(\gamma_1,0) [\tau(\gamma_2,0)-\tau(\gamma_1,0)] {\mathcal A}(\gamma_2,0)\\
        &&+[1+\mathcal{A}(\gamma_1,0)\tau(\gamma_1,0)]
        [\mathcal{Z}(\gamma_2,0)-\mathcal{Z}(\gamma_1,0)]
              [1+\tau(\gamma_2,0)\mathcal{A}(\gamma_2,0)]~.
\label{eq:gammareln}
\eea 
Once the threshold amplitude $\mathcal{A}(\gamma_2,0)$ is
obtained from this equation, it can in turn be used to calculate the
scattering amplitude at any energy and for the new value of $\gamma$, $\gamma_2$.

\begin{figure}[t]
\centerline{\includegraphics*[width=10cm,angle=0,clip=true]{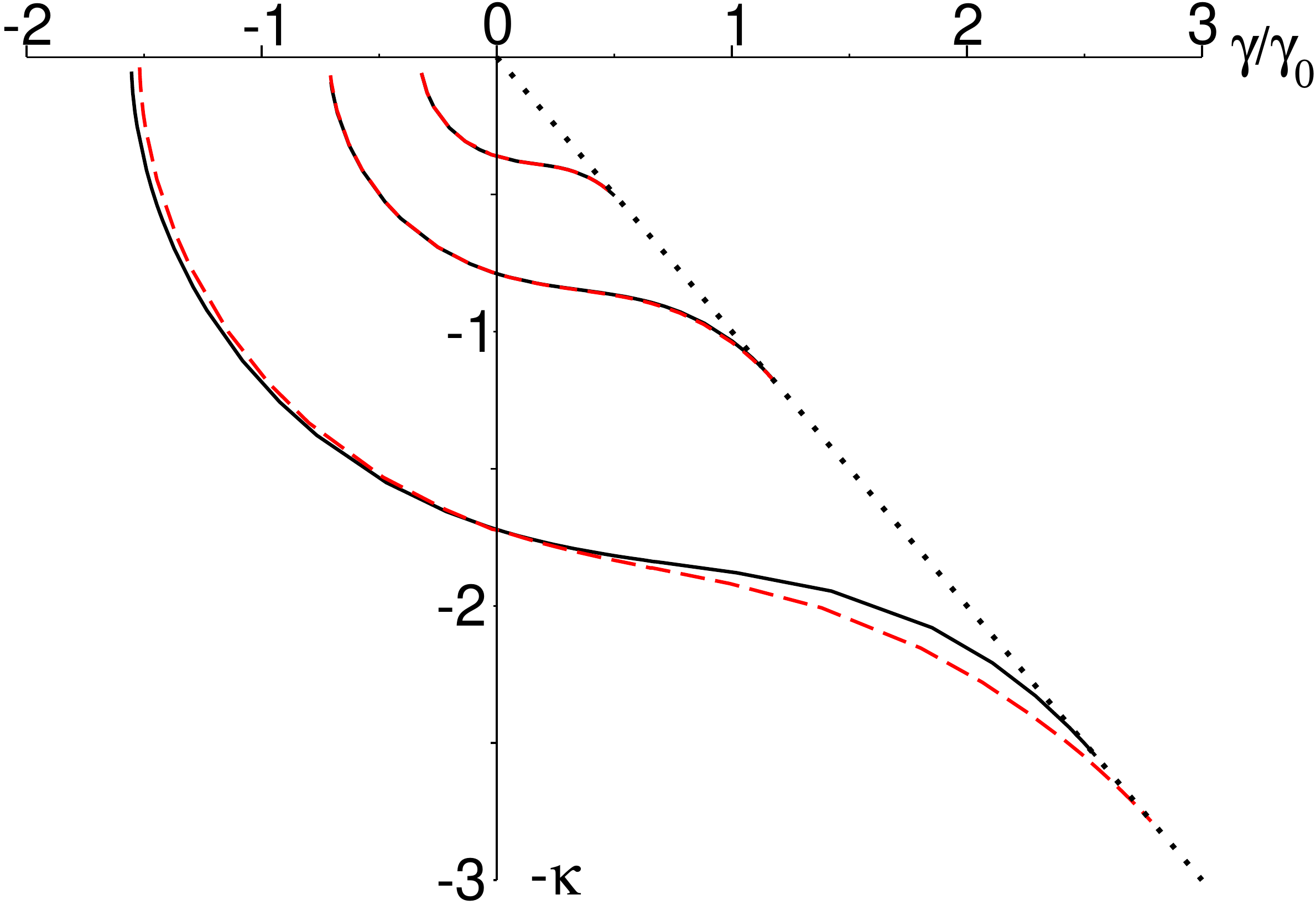}
}

\caption{The bound-state spectrum of the three-boson system with short-range interactions. The solid lines
denote the leading-order results and the dashed lines give the next-to-leading
order result for an effective range of $\gamma_0 r_s=0.01$. Here the scale has been
altered as per Ref.~\cite{Braaten:2004rn}, with $\gamma$ and $\kappa$ first
rewritten in polar form as $\gamma=\gamma_0 H\cos\zeta$ and $\kappa=\gamma_0 H\sin\zeta$, and then
the quantities $H^{1/4}\cos\zeta$ and  $-H^{1/4}\sin\zeta$ displayed here.
 The dotted line denotes the point at which trimers become unstable against
breakup to an atom and a dimer.}
\label{fig:efimovplot}
\end{figure}

The strategy is thus:
\begin{enumerate}
\item Pick a value of $a_3$ and set $\gamma_1=\gamma_0$ (chosen arbitrarily)
  and compute the threshold ${\cal A}$, ${\cal A}(\gamma_0,0)$, using
  Eqs.~(\ref{eq:threshsubt}) and (\ref{eq:fulloffshell}). 

\item Use Eq.~(\ref{eq:gammareln}) to compute ${\cal A}(0,0)$: the
  threshold amplitude in the unitary limit. 

\item Use Eq.~(\ref{eq:full-amplitude-subtracted}) to obtain ${\cal
  A}(\gamma=0)$ for $E<0$, thereby determining the bound-state spectrum.

\item Adjust $a_3(\gamma_0)$ and repeat steps 1-3 until a desired $\kappa_*$ is
  obtained. 

\item Then use Eq.~(\ref{eq:gammareln}) to compute ${\cal
  A}(\gamma,0)$ for other values of $\gamma$.

\item And once again use Eq.~(\ref{eq:full-amplitude-subtracted}) to
  determine the bound-state spectrum.
\end{enumerate}

The results of this procedure are shown in Fig.~\ref{fig:efimovplot}.
This represents the spectrum of three-body bound states which
correspond to different two-body scattering lengths but the same
asymptotic behavior in the short-distance region (i.e. the same
$\kappa_*$ at $\gamma=0$).  The binding energies are renormalized so that there is a state in the
unitary limit which has binding momentum $\kappa_*=\sqrt{0.15} \gamma_0$.
(From now on we denote this state as the one with index $n=1$.) We also set $M=\hbar=1$.
In Fig.~\ref{fig:efimovplot} the binding momentum
$\kappa=\sqrt{M B_3}/\hbar$ is plotted against the inverse of the two-body
scattering length. The solid line
represents the leading-order results, which reproduce those presented
in Ref.~\cite{Braaten:2004rn}. Although the axes are labelled $-\kappa$ and $\gamma$, we have
actually plotted $-H^{1/4} \sin\zeta$ versus $H^{1/4} \cos\zeta$, with (the dimensionless) $H$ and
$\zeta$ defined as $\gamma=\gamma_0 H\cos\zeta$ and $\kappa=\gamma_0 H\sin\zeta$. This
rescaling of the variables allows us to display a greater range $\gamma$ and
$\kappa$ as the discrete scaling factor is reduced from $22.7$ to
$22.7^{1/4}=2.2$~\cite{Braaten:2004rn}. 

We can extend these results to NLO by following Steps 1-7 above, but
using the form (\ref{eq:Sn}) for $n=1$ in the integral equations. We
reiterate that we renormalize at NLO by demanding that the value of
$\kappa_*$ is unchanged from its LO value. Consequently the value of
$a_3$ obtained at the starting point $\gamma=\gamma_0$ undergoes a
shift. In Fig.~\ref{fig:efimovplot} we show an extended {\it Efimov} plot
which includes the NLO results for a representative value of
$r_s \gamma_0=0.01$ as the dashed line. The effective range has been chosen such that the deepest three-body bound state shown remains
in the range of validity of the EFT for all allowed values of $\gamma$.
We see that the shift caused by range corrections is only
recognizable for this state, and that even the shift in its binding energy vanishes 
in the unitary limit. 

\subsection{Universality in the range correction}
\label{sec-smseuniv}

We now take the results and methodology of the previous subsection and
use them to compute corrections to the bound-state spectrum that are linear in the effective range $r_s$. This calculation
was performed in the limit $|a| \rightarrow \infty$ in Sec.~\ref{sec-linearcorrn}, and here we extend
it to the case
of finite $a$. In the small $r_s$ regime we can write (c.f. Eq.~(\ref{eq:Gn1}))
\begin{equation}
B_n=\frac{\hbar^2 \kappa_*^2}{M}\left[F_n\left(\frac{\gamma}{\gamma_0}\right) + \kappa_* r_s G_n\left(\frac{\gamma}{\gamma_0}\right) + \ldots\right]
\label{eq:Gdefn}
\end{equation}
where the function $F_n$ gives the leading-order binding energy of the $n$th state at arbitrary $\gamma$~\cite{Braaten:2004rn}. The scale
$\gamma_0$ is arbitrarily chosen, and for the calculations presented here we have $\kappa_*=\sqrt{0.15} \gamma_0$.
In general we will consider values of $\gamma$ which
range from $0$ (i.e. the unitary limit) to $\gamma \approx 14.2 \kappa_*$, where the $n=1$ trimer state vanishes through the atom-dimer threshold. It needs to be noted, though, that at any
non-zero $\gamma$ there are only a finite number of bound states within
the domain of validity of the EFT: the Efimov spectrum of infinitely
many shallow three-body bound states is only realized at
$\gamma=0$. Indeed, at $\gamma=\gamma_0$ there are only two three-body
bound states which are inside the domain of validity of the EFT.

\begin{figure}[t]
\centerline{
\includegraphics*[width=8.5cm,angle=0,clip=true]{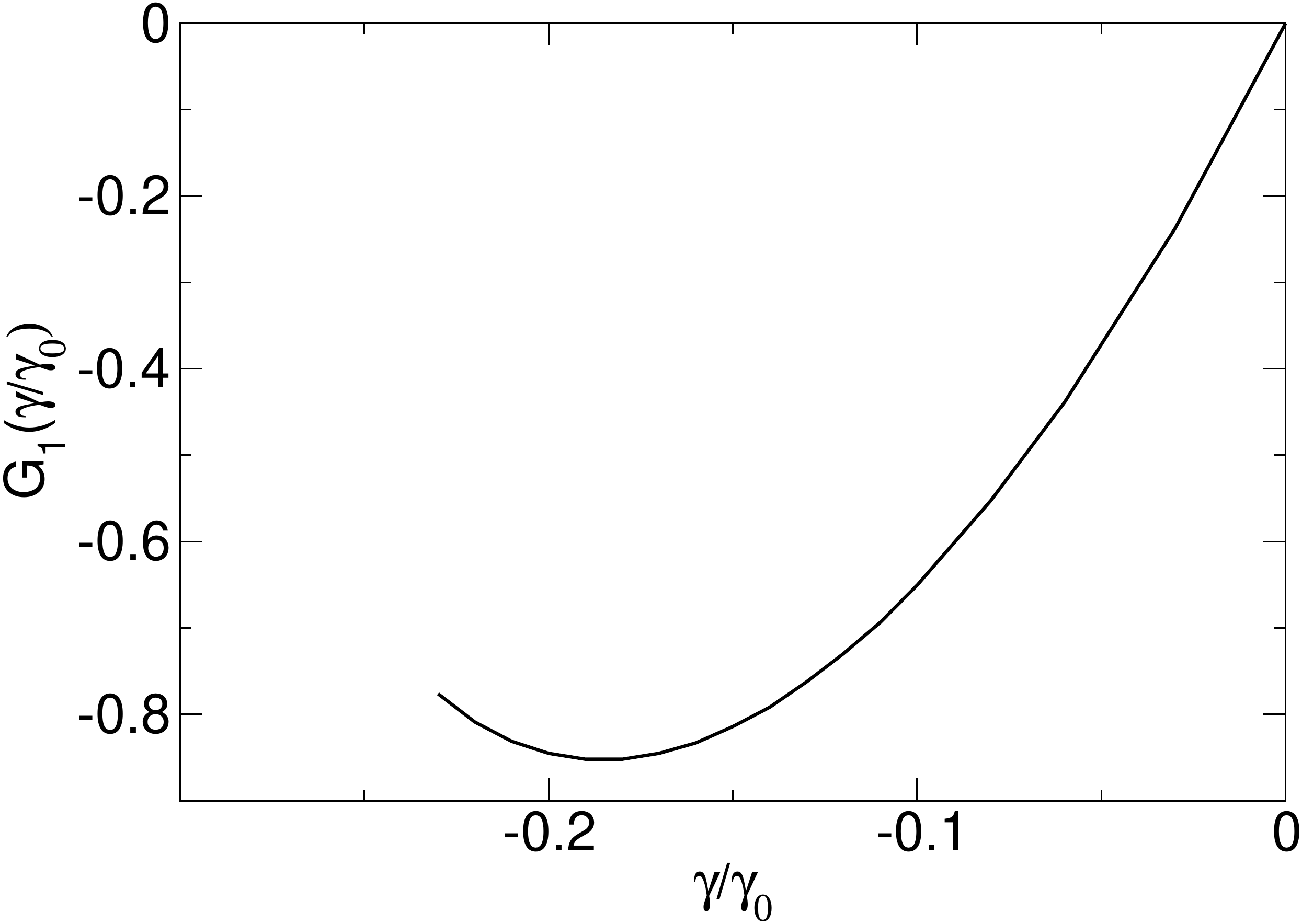}
\includegraphics*[width=8cm,angle=0,clip=true]{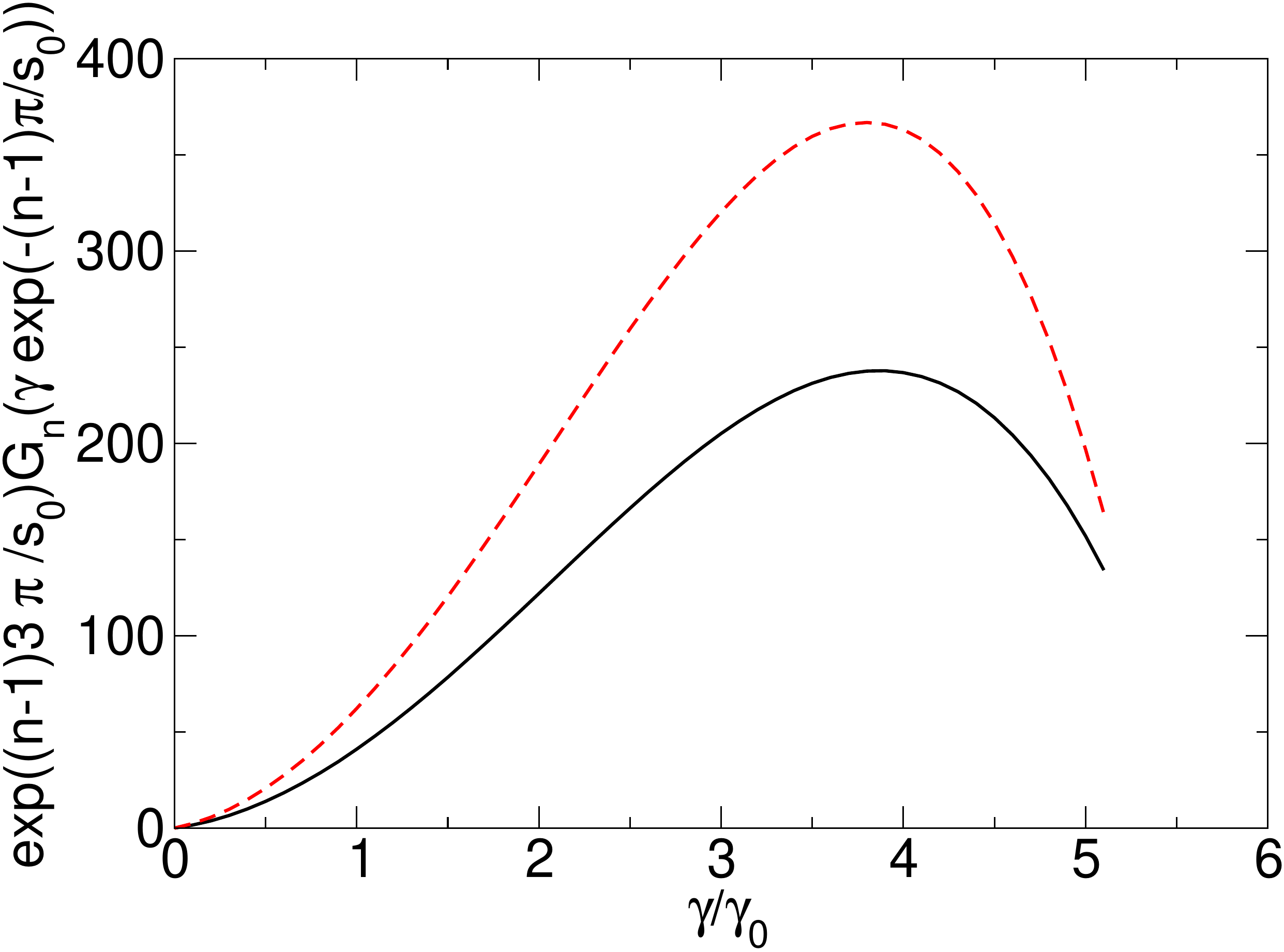}
}
\caption{Left panel: The extracted function $G_1$ for negative scattering
  length. Right panel: Results for $G_1$ and rescaled $G_2$ for positive scattering
  length The solid line denotes the result for the function $G$ extracted
from the bound state which has binding momentum $\kappa_*=\sqrt{0.15} \gamma_0^2$ in
the unitary limit. The dashed line denotes the rescaled result obtained
from the bound state which has binding momentum $\kappa=\kappa_*/22.7$
in the unitary limit.}
\label{fig-G1}
\end{figure}

Discrete scale invariance leads us to hypothesise that the $G_n$'s are related to $G_{n*}$ by:
\begin{equation}
G_n\left(\frac{\gamma}{\gamma_0}\right)=\exp\left(\frac{(n*-n)3\pi}{s_0}\right) 
\theta_n \left(\frac{\gamma}{\gamma_0}\right)
G_{n*} \left(\frac{\gamma}{\gamma_0} e^{\frac{(n-n*)\pi}{s_0}} \right),
\label{eq:Gn}
\end{equation}
where the factor out the front comes
from assuming $\kappa_n^3$ scaling of the first-order correction.
If we computed solely in the unitary limit this pre-factor would be
$e^{(n*-n) 3 \pi/s_0}$, but a number of effects that modify this scaling for finite $\gamma$
(when the hyperspherical potential is no longer a sum of power laws, even for small $r_s$) are accounted for 
by the function $\theta_n$. This obeys
\begin{eqnarray}
\theta_{n*}(z)&=&1~\mbox{for all $z$},\\
\theta_n(0)&=&1~\mbox{for all $n$}, \label{eq:constraint2}
\end{eqnarray}
and is $\sim 1$ everywhere else.
The constraint (\ref{eq:constraint2}) on $\theta_n$ 
incorporates the result already demonstrated analytically in the limit
$\gamma \rightarrow 0$: that if the linear-range correction vanishes
for one Efimov bound state then it vanishes for all of them.
This result is verified by both our results and an independent computation~\cite{Daekyoung}.

Here we choose $n*=1$.
The extraction of the function $G_n$ from Eq.~(\ref{eq:Gdefn}) is
numerically quite delicate. It involves a numerical derivative with
respect to $r_s$, but in order to keep consistency with the scales of
our EFT we must maintain both $r_s \gamma \ll 1$ and $\Lambda r_s \gg
1$. In practice we compute $G$ for a number of values of $r_s$ that
satisfy these constraints and then extrapolate to the $r_s \rightarrow
0$ limit.

The results for $G_1$ for positive $\gamma$ are shown as the solid line
in the right panel of Fig.~\ref{fig-G1}.
Away from $\gamma=0$
the case of $G_0$ does not give us a wide enough range of $\gamma$
over which to test Eq.~(\ref{eq:Gn}) while still remaining inside the
domain of validity of the EFT. 
We can, however, test the prediction (\ref{eq:Gn})  for the case
$n=2$. Since the binding energies for this case are two orders of
magnitude smaller, significantly higher accuracy is needed to compute
the necessary numerical derivatives.
Results for the function  $G_2$ in the case of positive two-body scattering length 
are represented by the dashed line in the right panel of Fig.~\ref{fig-G1}, with 
$G_2$ having been rescaled according to Eq.~(\ref{eq:Gn}). We see
that the horizontal scale in the two cases is indeed related by a factor of $e^{\pi/s_0}$, exactly
as predicted by discrete scale invariance. Also, both functions vanish at the same point, $\gamma \approx 5.5 \gamma_0$,
once this rescaling is performed ($\gamma \approx 0.24 \gamma_0$ for $n=2$ before rescaling). 
Moreover, the peak in the rescaled $G_2$ is at approximately the same position ($\gamma \approx 3.5 \gamma_0$)
as that in $G_1$. But, although the factor of $22.7^3$ postulated in Eq.~(\ref{eq:Gn}) allows functions that differ by four
orders of magnitude to be plotted on the same vertical scale, the peak of $G_2$ after rescaling is 30\% higher than 
anticipated based on $G_1$ and the scaling of binding energies in the
unitary limit. Further investigation of these deviations from pure $\kappa_n^3$ scaling that are encoded in 
the functions $\theta_n$ will be interesting, but are beyond the scope of this paper.

That the rescaled $G_2$ and the function $G_1$ vanish at the same value of $\gamma$ would seem 
to be a simple
consequence of the fact that the binding momentum of the three-body
state goes to zero at this point, and so the first-order correction to the
binding energy must be zero.  But this
straightforward observation has the important consequence that corrections linear in $r_s$ do not
affect the values of
$\gamma > 0$ at which states that would be part of the Efimov tower in the
$\gamma=0$ limit disappear from the bound-state spectrum. 
We conjecture that this result also applies to the case of negative scattering length,
i.e. $\gamma < 0$.

In the left panel of Fig.~\ref{fig-G1} we display the result
for $G_1$ at negative scattering length. The rapid decrease of the
three-body binding energies in this domain
makes the extraction of the
functions still more complicated, and in this case we achieve 
satisfactory numerical accuracy only for $G_1$. The state with $n=1$ vanishes
above the three-body threshold at $\gamma=-0.26(1) \gamma_0$, so we obtain $G_1$
over almost the entire range where it is defined.
The particularly fast variation of three-body binding energies near this vanishing point means that we could not extract the value of $G_1$ at the endpoint with any serious precision. Thus our conjecture of the previous paragraph remains unconfirmed. But, if true, that conjecture has important implications for three-body
recombination at negative scattering length. We therefore now turn our attention to how recombination can be computed
from the scattering amplitudes we have already obtained.


\section{Range Corrections to Three-Body Recombination}
\label{sec-recombination}
In this section we will consider effective-range corrections to the
three-body recombination rate for positive two-body scattering length.
The calculation of the rate at negative scattering length requires the
evaluation of the S-matrix element for three-atom-to-three-atom scattering.
This is more involved and will be considered in a later publication.

Three-body recombination is a process in which three atoms
collide to form a diatomic bound state and the energy that is thereby released causes the
particles  to leave the trap. The rate of decrease in the number density of
atoms in the trap is given by
\beq
\frac{\hbox{d}}{\hbox{d}t}n=-3\alpha n^3~.
\eeq
The factor of $3$ in the above equation arises from the assumption that
all three particles involved in the recombination process will leave the
trap. 
Atoms with large positive scattering length can recombine into the shallow
dimer with $E_D=\hbar^2 \gamma^2/M$. Recombination can also produce deep dimers with $E_{\rm deep}\sim
\hbar^2/(M l^2)$---if such states are supported by the underlying two-body interaction. For
negative scattering length only recombination into such deep dimers is possible.

Range corrections to three-body recombination have previously been considered
in an EFT framework~\cite{Hammer:2006zs}. But there
only correlations between the atom-dimer scattering length and
$\alpha$ were considered. However, the effect that range
corrections have on $\alpha$ as the two-body scattering length is varied but the three-body
parameter (here denoted by $\kappa_*$) remains fixed is particularly relevant to experiment. This case corresponds to the situation
in which the two-atom scattering length is tuned using an external
magnetic field. Sizeable sensitivity of $\gamma$ (or equivalently $a$) to the magnetic field is called a Feshbach resonance
and occurs if a change in the magnetic-field strength leads to a threshold crossing of a two-body
bound state.

\begin{figure}
\centerline{\includegraphics*[width=10cm,angle=0,clip=true]{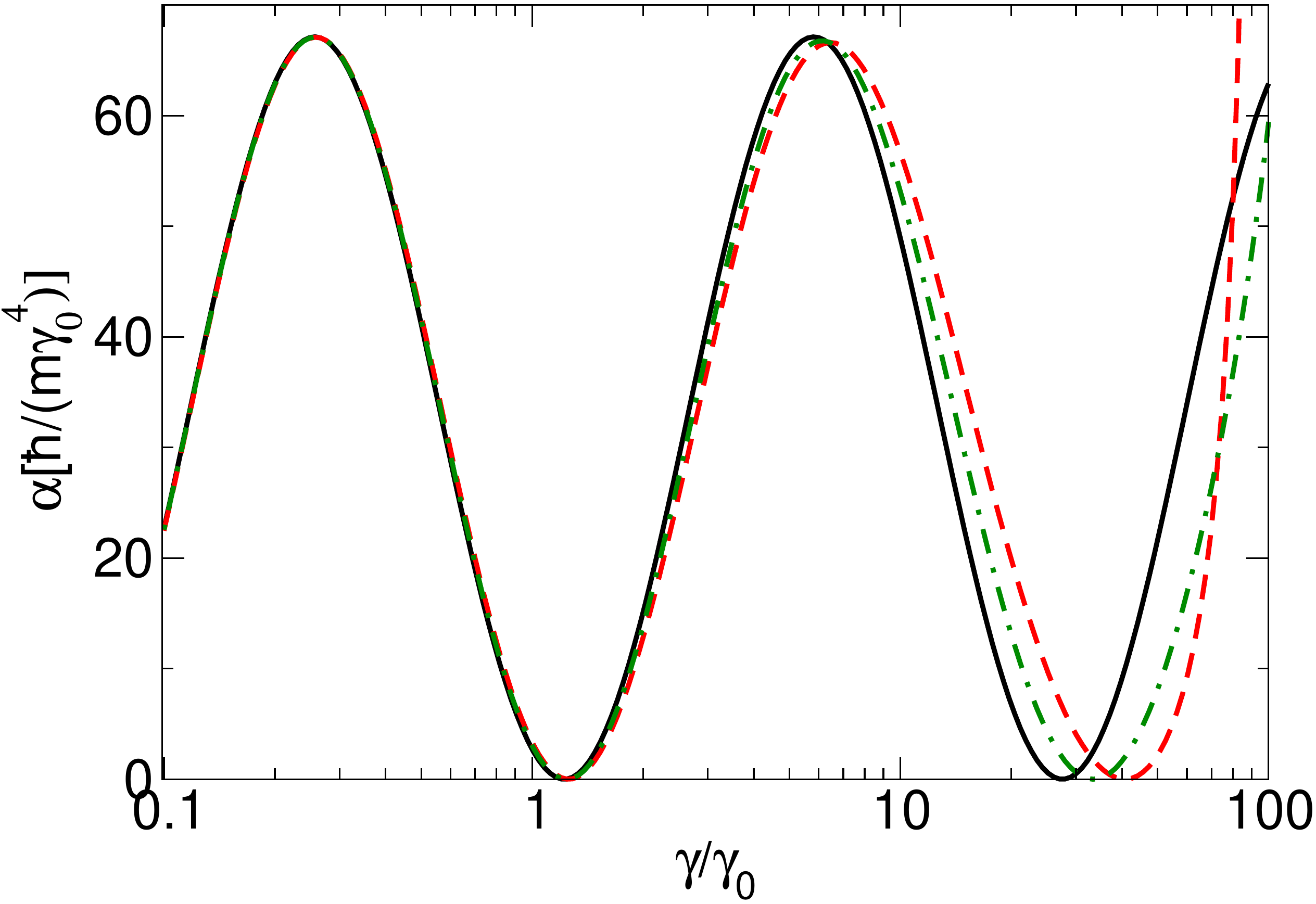}
}
\caption{The recombination rate $\alpha$ as a function of $\gamma/\gamma_0$.
The solid line gives the leading order result, while the dashed and dot-dashed
lines give the results for $r_s \gamma_0=0.01$ and $0.005$, respectively.}
\label{fig:recombination}
\end{figure}
Here, for simplicity, we consider three-body recombination at threshold.
This corresponds to the zero-temperature limit. We will also
neglect three-body recombination into deep dimers. An analytic form for the
leading order recombination rate into shallow dimers at threshold has been
derived by Macek {\it et al.} \cite{MOG05} and independently by
Petrov \cite{Petrov-octs}
\begin{eqnarray}
\alpha(E=0) = 
\frac{128 \pi^2 (4 \pi - 3 \sqrt{3}) \sin^2 (s_0 \ln\frac{a}{a_{*0}})}
    {\sinh^2(\pi s_0) + \cos^2 (s_0 \ln\frac{a}{a_{*0}})} \,
\frac{\hbar}{M \gamma^4} \,.
\label{K0-analytic}
\end{eqnarray}
The recombination rate therefore scales as the product of $\hbar/(M \gamma^4)$ and
a log-periodic function.
The three-body parameter $a_{*0}$ that determines the phase of this log periodicity defines the
scattering length at which the recombination rate has a minimum. At leading order it is related to the three-body parameter $\kappa_*$ by $a_{*0}=(e^{\pi s_0})^n 0.32 \kappa_*^{-1}$~\cite{Braaten:2004rn}.

The recombination rate can be obtained from the elastic atom-dimer
scattering amplitude ${\cal A}$ by solving the STM equation and evaluating
\cite{Bedaque:2000ft,Hammer:2006zs}
\beq
\label{eq:reco}
\alpha=\frac{64\pi^2 M}{\sqrt{3}}\frac{1}{(1-\gamma\frac{r_s}{2})^2(1-\gamma
  r)}|\mathcal{A}(0,2\gamma/\sqrt{3},0)|^2~.
\eeq
At leading order this reproduces Eq.~(\ref{K0-analytic}). We now use the subtracted integral
equation derived in Sec.~\ref{sec-pspace} to calculate the amplitude $\mathcal{A}$ in Eq.~(\ref{eq:reco}).
In any experiment the effective range will vary with
the external magnetic field which is used to control the scattering
length. However, this variation of $r_s$ will be slow compared to that of $\gamma$, and so
we assume that the effective range stays constant as $\gamma$ varies.

Figure \ref{fig:recombination} shows the resulting dependence of the threshold recombination rate $\alpha$
on $\gamma$ in units of $\hbar/(\gamma_0^4 M)$.
The results display the  log-periodic behavior expected from Eq.~(\ref{eq:reco}).
The solid line gives the three-body recombination
rate in the zero-range limit---the LO EFT result---while the dashed and dot-dashed lines give the
NLO results for $r_s \gamma_0=0.01$ and $0.005$, respectively.
All calculations are renormalized so that they correspond to the same binding momentum $\kappa_*$
of the first excited Efimov state at $|a|=\infty$.  The renormalization at $|a|=\infty$
makes all curves approach the LO result as $\gamma \rightarrow 0$.
Away from $\gamma=0$ it can be seen that the finite effective
range leads to a shift of the minima in the recombination rate and also influences the
maximum value of $\alpha$. 

Given
the recent experiment of Ref.~\cite{Grimm06}
the extension of this computation to the case $\gamma < 0$ will be particularly interesting. In particular, the minima in the recombination
rate are determined by the values of the scattering length at which one of
the Efimov states is at threshold. Therefore we conjecture that these minima {\it receive no 
correction of $O(r_s)$}. This will be discussed in a future publication. More generally for both $\gamma > 0$ and $\gamma <  0$ the inclusion of range corrections into calculations for the
recombination of $^{133}$Cs atoms will allow for a better description
of data and therefore a more precise determination of the relevant
three-body parameters. However, in order to obtain truly precise results it will be necessary to account both
for the effects of deep dimers and the variation of the effective range as
a function of the magnetic field.

\section{Summary and Outlook}
\label{sec-conclusion}

Near a Feshbach resonance the scattering length, $a$, of the two-body
system is much larger than all other length scales in the problem. This
leads to an Efimov spectrum in the three-body system, which displays
discrete scale invariance.  However, corrections due to the effective
range, $r_s$, will in principle always affect observables even in the limit of
infinite scattering length. This paper
considered the impact of such corrections on three-body system
observables.

We did this first by examining the spectrum of the three-body system
in the unitary limit, where $|a| \rightarrow \infty$. Using
hyperspherical coordinates for the three-body system, we extended
previous analyses~\cite{FedorovJensen,Braaten:2004rn} to derive the
potential that corrects the strict $1/R^2$ potential present in the
unitary limit for the presence of a non-zero $r_s$. (This potential
had been conjectured, but not derived in Ref.~\cite{Efimov91}.)
Specifically, we found that the perturbing potential is
\begin{equation}
V^{(1)}(R)=-\frac{s_0^2\, \xi_0 \,r_s}{R^3},
\label{eq:V1}
\end{equation}
where $s_0\approx 1.00624$ is the eigenvalue of the leading-order
hyper-angular equation and $\xi_0 \approx 0.480$ is obtained by analyzing that
equation for $r_s \neq 0$.

We assessed the impact of the potential $V^{(1)}$ on the Efimov spectrum of
three-body bound states. Since the potential is singular renormalization is required. 
If we take as our renormalization
condition that the $n$*th state is unperturbed by the $V^{(1)}$
correction, then the shift in the $n$th state is zero in the limit $\Lambda
\rightarrow \infty$, and so the entire spectrum is unperturbed. This is
consistent with the discrete scale invariance that is present in that
limit, but we note that the conclusion is dependent on the particular
renormalization condition chosen. 

We then examined how these results change as we move away from the
limit $|a| \rightarrow \infty$. We derived subtracted integral
equations that allowed us to relate the $1+2$ scattering amplitude at
different values of the scattering length $a$. This led us to an
extended version of the Efimov plot, where we were able to display how
three-body bound state energies vary not only with $a$, but also
with $r_s$. Such a plot could also be generated with an explicit
three-body force, but the subtractive approach employed here has
certain advantages, in particular when higher-order calculations at
large cutoffs are pursued. This approach has been used with success in 
Refs.~\cite{Platter:2006a,Platter:2006b,Hammer:2006zs,Hammer:2007a}
in a variety of nuclear and atomic-physics contexts to compute results that are
claimed to be valid up to N2LO in the $r_s/a$ expansion. It is important to 
point out that the disagreement between these works and Refs.~\cite{Bedaque:2002yg,Barford:2004fz,Griesshammer:2004pe,Griesshammer:2005}
regarding whether additional three-body input is needed to renormalize the three-body problem
at N2LO does not affect the results presented here. The formalism
employed to compute the first-order correction in $r_s$ reproduces the results
of Refs.~\cite{Bedaque:2002yg,Barford:2004fz,Griesshammer:2004pe,Griesshammer:2005}, even if it leads
to conclusions at N2LO that disagree with these studies. 

Thus the only assumption in our work is that $|r_s| \gg l$, where $l$ is the underlying length scale
in the two-body potential. Because of this assumption all calculations of Sec.~\ref{sec-pspace} were performed
in the limit $\Lambda \rightarrow \infty$.We extracted, in this limit, a universal
function that describes the linear (in $r_s$) correction to the
bound-state energies of the three-body bound-state spectrum.  We chose to examine the correction
for the state with $n=n*=1$, with the universal function defined by:
\beq B_{n*}=\frac{\hbar^2 \kappa_*^2}{M}
\left[F_{n*} \left(\frac{\gamma}{\kappa_*}\right) + \kappa_* \,r_s
  \,G_{n*}\left(\frac{\gamma}{\kappa_*}\right) + O[(\kappa_*\,
    r_s)^2]\right], \eeq where $F_n$ is given in Ref.~\cite{Braaten:2004rn},
and defines the $\gamma$ dependence of the bound state in the $r_s=0$
(scaling) limit. The result we obtained for $G_{n*}$ is displayed in Fig.~\ref{fig-G1}.
We have furthermore showed that the next-to-leading order corrections
to different states in the Efimov spectrum are approximately related to each other
by the scale transformation defined in Eq.~(\ref{eq:Gn}).
It is an interesting question whether this symmetry of the 
leading-order wave function impacts other observables too.

Being able to use subtracted equations to relate physical observables
at different scattering lengths to each other led us to consider
three-body recombination into the shallow dimer. We calculated
the recombination rate for different values of the effective range
$r_s$ while renormalizing to the binding momentum $\kappa_*$ of
an Efimov state in the unitary limit. Consequently our results
for different values of the two-body effective range approach one another
as the two-body scattering length increases.

In order to examine the behavior of the three-body system near a
Feshbach resonance we have used an EFT that exploits the hierarchy of
scales $|r_s| \ll |a|$. We have focused on effects of $O(r_s/a)$. This
enables us to probe the pattern of convergence of the EFT, thereby
facilitating reliable error estimates for observables calculated in
this framework. 
It also provides a clear differentiation between effects that
are universal (i.e. arise solely as a result of $|a| \gg |r_s|$) and effects
that depend on details of the underlying two-body interaction.
It is interesting to speculate that features such as the
Phillips and Tjon lines in nuclear physics are well reproduced at
leading order~\cite{Bedaque:1999ve,Platter:2004zs}, in spite of the large expansion parameter there ($r_s/a
\approx 0.3$) because the discrete scale invariance suppresses the NLO
correction below the natural expectation.

However, in order to confirm this speculation an analysis of the
four-body system in, and near, the unitary limit must be carried
out. Such a conjecture might also be rendered more plausible if the
discrete scale invariance that is present at leading order suppressed
some of the next-to-next-to-leading order corrections to the Efimov
spectrum. In order to examine this possibility we would like to extend
the above analysis to $O(r_s^2/a^2)$.  While the computation of such
corrections to the bound-state spectrum appears straightforward, it is
numerically somewhat delicate. However, a computation of the
$O(r_s^2)$ corrections to the unitary-limit spectrum can be carried
out using the hyperradial formalism in a manner similar to that
employed here.

\begin{acknowledgments}
This work was supported by the Department of Energy under
grant DE-FG02-93ER40756, by the National Science 
Foundation under Grant No. PHY--0653312 (LP),
the UNEDF SciDAC Collaboration under DOE Grant 
DE-FC02-07ER41457 (LP) and by an Ohio University postdoctoral
fellowship (LP). We thank Daekyoung Kang for generating the data that
allowed us to check Eq.~(\ref{eq:DeltaBnfinal}), and Eric Braaten for 
stimulating this study by drawing our attention to Ref.~\cite{Efimov91}. DRP is grateful to the 
Theoretical Physics group at the University of Manchester for hospitality
during the completion of this work. 
\end{acknowledgments}
\begin{appendix}
\section{A problem with $r_s \gamma < 0$}
For $r_s \gamma < 0$ problems with the method we have adopted to
obtain the correction $\sim r_s$ in the bound-state spectrum occur as
soon as we consider cutoffs $\Lambda \gg 1/r$. In this case the
largest eigenvalues of the kernel of the original STM equation
(\ref{eq:stm})
are negative, and so the subtraction does not improve the equation's
behavior. Instead, when we solve the subtracted integral equation 
(\ref{eq:full-amplitude-subtracted})
at NLO we obtain spurious low-energy solutions which have no LO
counterpart, i.e. they are not smoothly connected to a LO eigenvector
by variation of $r_s$. Examining these eigenvectors we see that they
have most of their support at $p \sim 1/r$. These eigenvectors are
thus a non-perturbative effect generated by the inclusion of the
pieces $\sim r_s$ in the kernel. The effect of these NLO pieces of the
kernel should be perturbatively small, so the fact that they produce
eigenvectors with support at $p \sim 1/r$ is not a valid prediction of
the EFT.

Therefore we wish to eliminate these eigenvectors from the spectrum of
the kernel, leaving us with only eigenvectors (and eigenvalues) which
are perturbatively close to those that exist at leading order. We can
do this by noting that, regardless of the sign of $r_s$, the integral
equation with the N2LO kernel above yields small changes from the LO
result for small $r_s$. The largest eigenvalues of the N2LO kernel
($n=2$ in Eq.~(\ref{eq:Sn}))
are positive regardless of the sign of $r_s$---as in the LO case. Thus
the subtraction approach works straightforwardly there.  Therefore,
we choose to stabilize the troublesome NLO integral equation for NLO
by adding an admixture of the N2LO kernel to the NLO one. In other
words, we choose for the residue function $S(E;q)$
\begin{eqnarray}
S_\alpha^{(1)}(E;p) \equiv S^{(1)}_\alpha \left (E-\frac{3 \hbar^2 p^2}{4M}\right)&=&
\frac{2 \hbar^4}{\pi  M^2}\left\{\left(\gamma+\sqrt{\frac{3}{4} p^2 -
  \frac{mE}{\hbar^2}}\right)
(1 + \gamma r_s) + \frac{3}{8} r_s (p^2 - k^2) \right.\nonumber\\
&\quad&\qquad \quad \left.-\alpha \frac{3 r_s^2}{16} (p^2 - k^2) \left(\gamma - \sqrt{\frac{3}{4} p^2 - \frac{ME}{\hbar^2}}\right)\right\},
\label{eq:Salpha}
\end{eqnarray}
with $E$ and $k$ related by $E=\frac{3}{4} \frac{\hbar^2 k^2}{M} - E_D$.
Here the piece on the second line effects the stabilization, but note
that it is not the full N2LO correction, since we have omitted pieces of the
kernel that scale as $(\gamma r_s)^2$, as they do not assist with
the stabilization. Therefore $S_1^{(1)} \neq S^{(2)}$. This distinction
is irrelevant in the unitary limit, and so there (\ref{eq:Salpha})
interpolates smoothly between the NLO result (for $\alpha=0$) and the
N2LO result for ($\alpha=1$). And, provided perturbation theory applies,
the result of using $S^{(1)}_\alpha$ instead of $S^{(1)}$ should be
linear in $\alpha$. Therefore as long as $r_s$ is small enough that
perturbation theory is valid we can examine the results for
calculations with $S^{(1+\alpha)}$ and make a linear extrapolation to
$\alpha=0$ in order to obtain the NLO result.

Below we show a table of results for $B_3^{(1)}$ using
$S^{(1)}_\alpha$ for $r_s \gamma=-0.07$, and $r_s \gamma=-0.14$ and
$a_3 \gamma=1.79$. Once $\alpha < 0.3$ the admixture of the N2LO piece
in the propagator no longer stabilizes the integral equation. However,
the data for $\alpha > 0.4$ is already sufficient to allow us to
predict:
\begin{equation}
B_3^{(1)}/E_D (r_s \gamma=-0.07)=1.7115(1),
\end{equation}
and 
\begin{equation}
B_3^{(1)}/E_D (r_s \gamma=-0.14)=1.6983(1),
\end{equation}
for the pure NLO piece of the result. When combined with the LO
result:
\begin{equation}
B_3^{(1)}/E_D (r_s \gamma=0.0)=1.723,
\end{equation}
this provides nice evidence of linearity in $r_s \gamma$, even though
we are now considering
$r_s < 0$.
Moreover, these predictions are entirely consistent with the NLO results for
positive $r_s \gamma$:
\begin{equation}
B_3^{(1)}/E_D (r_s \gamma=0.07)=1.734; \qquad
B_3^{(1)}/E_D (r_s \gamma=0.14)=1.743,
\end{equation}
and the assumption of linearity in $r_s \gamma$.

\begin{table}[tbp]
\begin{center}
\begin{tabular}{|c|c|c|}
\hline\hline
$\alpha$ & $B_3^{(1)}/E_D (r_s \gamma=-0.07)$ & $B_3^{(1)}/E_D 
(r_s \gamma=-0.14)$ \\\hline\hline
0.1 & 4.735 & 2.5761\\
0.2 & 1.7132 & 1.7021\\
0.25 & 1.7122 & 1.7000\\
0.3 & 1.7121 & 1.6995\\
0.35 & 1.7120 & 1.6993\\
0.4 & 1.7121 & 1.6993\\
0.45 & 1.7121 & 1.6993\\
0.5 & 1.7121 & 1.6993\\
0.6 & 1.7122 & 1.6995\\
0.75 & 1.7124 & 1.6998\\
1.0 & 1.7127 & 1.7004\\
\hline
\end{tabular}
\vspace{0.3cm}
\end{center}
\caption{$B_3^{(1)}$ as a function of the stabilization parameter
  $\alpha$. All results are accurate to the number of figures quoted.}
\label{table-1}
\end{table}

Indeed, ultimately this is the solution to the instability we are
attempting to cure: if perturbation theory is valid then the shift at
NLO should be linear in $r_s$. Since we can compute it for positive
$r_s$ using the methods outlined above we can extrapolate those
results linearly to the region $r_s < 0$ and thereby obtain the
perturbative shift in bound-state energies. If that result is
different from what is obtained by the use of the NLO kernel with $r_s
< 0$ then that difference is an effect beyond perturbation theory in
$r_s \gamma$ and $r_s \kappa$, and such effects are not our concern
here.
\end{appendix}
 
\end{document}